\documentclass[12pt]{article}
\usepackage{amsmath,amssymb,amsfonts,latexsym,epsfig,color}
\begin{document}
\title{A Scaling Law for Quark Masses \\ and the CKM matrix}
\author{Alp Deniz \"Ozer \thanks{oezer@theorie.physik.uni-muenchen.de}   \\
{\small Ludwig Maximillians University,   }  \\
{\small Physics Section , Theresienstr. 37, 80333  M\"unich
Germany}}
\date{}
\maketitle
\abstract{We have recently argued that quark masses may follow
a simple scaling law.  In this paper we  build
a simple mass matrix for quarks that can reproduce the scaling law
expression. The simple mass matrices of the model are then
generalized through general rotations in the flavor space,
including phase transformations. In turn they will be used to
construct the quark-mixing matrix. It has been found that the
model can predict the entries of the $CKM$ matrix in excellent
agreement with current values. We give precise values for the
light quark masses and determine the magnitude of the $CP$
violation and also the quark-mixing angles in the flavor space.
The main motivation behind this work is to relate the scaling law
predictions with quark-mixing, through a simple mass matrix and
its generalized Hermitian form.}
\section{Introduction}\label{sec:intro}

Many of the observables  in particle physics are related with
broken symmetries. In this respect, the masses of the quarks could
either be related with a Yukawa term, where the masses are due to
a higgs scalar field coupling to fermions, or they can result as a
departure from a chiral symmetry in the qcd sector. A remarkable
feature of the quark masses is that they exhibit a clear
hierarchy. We had discussed in a previous work ~\cite{Alp} that
the quark masses might follow a simple scaling law;
\begin{equation}\label{eq:scaling-law}
\begin{split}
& \frac{m_t}{m_c}=\epsilon^2_u \frac{m_c}{m_u}      \\
\end{split} \ , \ \ \ \ \ \ \
\begin{split}
& \frac{m_b}{m_s}=\epsilon^2_d \frac{m_s}{m_d}   \\
\end{split}
\end{equation}
where we had initially considered the case $\epsilon_u=\epsilon_d=1$.
It is well known that the u-type quark masses  consistently
satisfy the scaling expression for $\epsilon_u=1$. Unfortunately
the scaling expression does not properly accommodate\footnote{one
can not find a strange quark mass and a down quark mass that
satisfies the predictions of the current algebra among light quark
masses.} the d-type quark masses for $\epsilon_d=1$. If these
scaling expressions are to make any sense at all, it is necessary
to look for values of $\epsilon_d$ other than 1.

\section{Scaling U-type Q-masses ($\epsilon_u=1$)}\label{sec:sqm-utype}

Let us first demonstrate that the u-type quark masses satisfy the
scaling law expression for $\epsilon_u=1$. In this respect, we
will use the currently known values of u-type quark masses. From
the other side the simple scaling law makes sense only, if the
quark masses are all renormalized at the same energy scale.
Therefore for the u-type-quarks we choose the central value of
$m_t=174.3 \pm 5.1 $ GeV as a useful scale. We choose for the
c-quark mass $m_{c}(m_c)=1.27 \pm 0.05 $ GeV given
in~\cite{Leut-h}, which rescales to $m_c{(m_t)}=0.59$ GeV to
$0.65$ GeV, using the QCD renormalization group~\cite{Muta} with
$\Lambda=211^{+34}_{-30}$ MeV for five flavors ~\cite{Bethke}. The
u-mass $m_{u}$ is given as $m_{u}(1\, \text{GeV})=5.1 \pm 0.9 $
MeV~\cite{Leut-l}. Using the QCD renormalization group with
$\Lambda=211^{+34}_{-30}$ MeV for five flavors, one has
$m_{u}(m_{t})=1.87$ MeV to $2.68$ MeV.

A graphical approach based on the above summarized u-type quark
masses would be very useful for the demonstration.
\begin{figure}[thb]
\begin{center}
\includegraphics[scale=0.9]{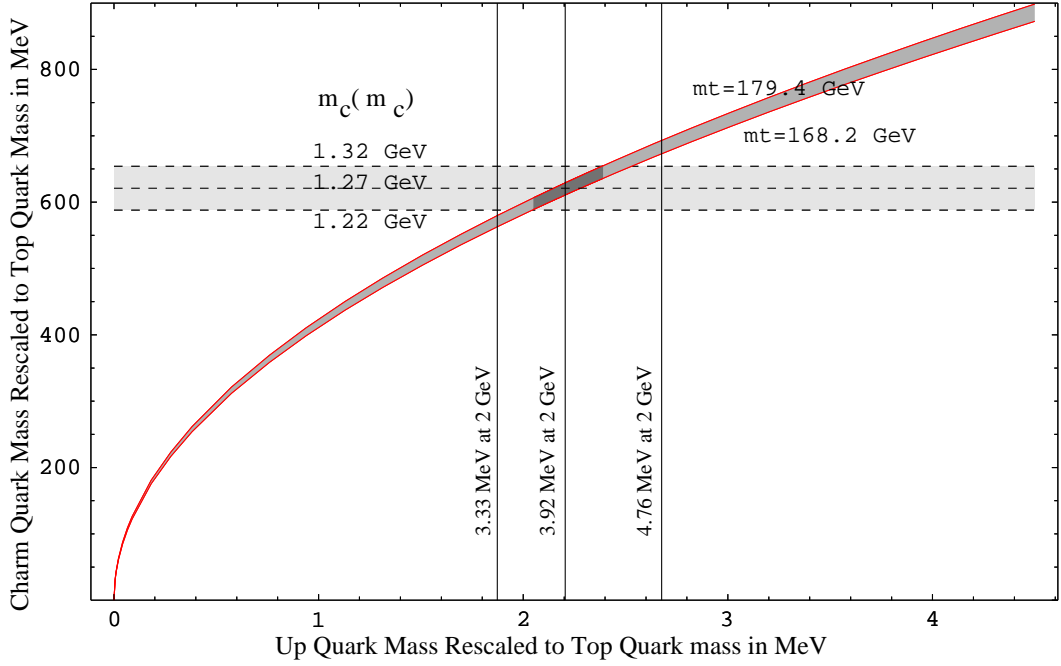}
\end{center} \caption{
U-type quark masses falling into the dark region between the two
curves satisfy the scaling law and the current bounds of u-type
quark masses. Note that $m_c$ and $m_u$ axis are rescaled to
$m_t=174.3$ GeV} \label{fig:up-quark}
\end{figure}
Using the scaling law given in  Eq.(\ref{eq:scaling-law}) we
obtain for the charm quark mass, $m_c=\sqrt{m_t \, m_u
}/\epsilon_u$. Since the top quark mass is known with a relatively
good precision we can plot a relation between $m_c$ and $m_u$,
once for the highest and once for the lowest values in $m_t=174.3
\pm 5.1$ GeV. The plot is shown in Fig. (\ref{fig:up-quark}).
Along the  curves the top quark mass is kept constant and the two
curves correspond to the upper and lower limit in the t-quark mass
which differ by the amount of $2*5.1$ GeV. The values of  $m_c$
and $m_u$  falling into the region between the two,  do
automatically satisfy the scaling law. The current value of the
c-quark mass from~\cite{Leut-h} is marked in the graphic as a grey
stripe which lies between the dashed lines. The upper and lower
dashed lines correspond to the highest and lowest values in
$m_c{(m_c)}=1.27 \pm 0.05$ GeV. The intermediate dashed line in
the grey region corresponds to the central value of
$m_c{(m_c)}=1.27$ GeV. These are the running masses in the
$\overline{MS}$ scheme and the highest and lowest values are
rescaled to $m_c{(m_t)}= 588$ MeV to $654$ MeV respectively. The
current best value of u-quark masses given in~\cite{Leut-l}
rescales to $m_{u}(m_{t})=1.87$ MeV to $2.68$ MeV and  are marked
in the graphic by the two vertical lines. These masses correspond
to $m_u=3.33$ MeV and $4.76$ MeV at $2$ GeV as indicated in the
figure. The vertical line intermediate to the other 2 vertical
lines corresponds to $m_u=3.93$ MeV at 2 GeV. The darkest region
which is the intersection of all the three regions, shows then the
u-type quark masses which satisfy scaling law and which do
simultaneously fall in the current limits of the u-type quark
mass.  The middle vertical and middle dashed line  are quite well
centered values with respect to  the dark region.

%%%%%%%%%%%%%%%%%%%%%%%%%%%%%%%%%%%%%
 %%%%%%%%%%%%%%%%%%%%%%%%%%%%%%%%%%%%%
 %%%%%%%%%%%%%%%%%%%%%%%%%%%%%%%%%%%%%
 %%%%%%%%%%%%%%%%%%%%%%%%%%%%%%%%%%%%%
 %%%%%%%%%%%DDDDDDDDDDDDDDDD%%%%%%%%%%
 %%%%%%%%%%%%%%%%%%%%%%%%%%%%%%%%%%%%%
 %%%%%%%%%%%%%%%%%%%%%%%%%%%%%%%%%%%%%
 %%%%%%%%%%%%%%%%%%%%%%%%%%%%%%%%%%%%%
\section{Scaling  D-type Q-masses ($\epsilon_d = \sqrt{2} $)}\label{sec:sqm-dtype}

The same can be done for the d-type quarks. Among the d-type
quarks $only$ the bottom quark is a "heavy quark" and has a
relatively well known mass. Unlike to the scale used in the former section ,the scaling
of the d-type quarks will be done at $2$ GeV. For the bottom quark
we choose $m_b(m_b)=4.25 \pm 0.10$ \cite{Leut-h} which rescales to
$m_b(2\,\text{GeV})=4.88$ GeV to $5.15$ GeV by using the QCD
renormalization group with the current value
$\Lambda=294^{+42}_{-38}$ MeV, for 4 flavors given in "$\alpha_s$
2002" by Bethke~\cite{Bethke}.  We chose for the strange quark
mass $m_s(1 \,\text{GeV}) = 175 \pm 25 $ MeV \cite{Leut-l}, which
rescales to $m_s=114$ MeV to $153$ MeV at $2$ GeV by using the QCD
renormalization group with the current value
$\Lambda=336^{+42}_{-38}$ MeV, for 3 flavors given in
~\cite{Bethke}. The down quark mass is chosen as $m_d(  \text{1
GeV})=9.3 \pm 1.4$ MeV \cite{Leut-l} which rescales to $m_d =
6.03$ MeV to $8.17 $ MeV.
\begin{figure}[thb]
\begin{center}
\includegraphics[scale=0.9]{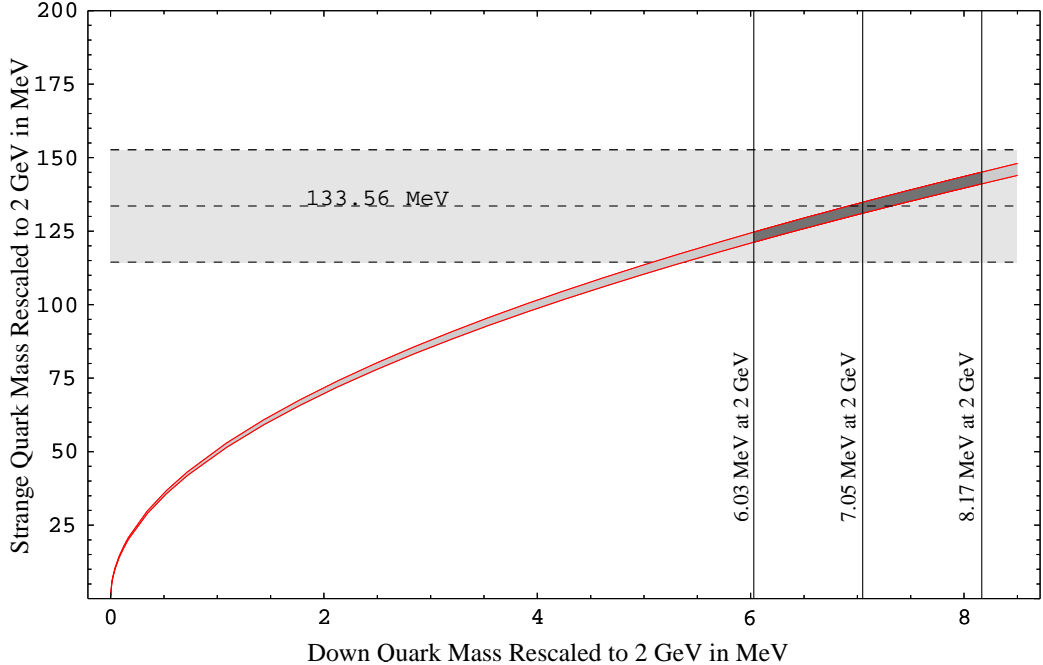}
\end{center} \caption{D-type quark masses falling into the dark  region between the two  curves
satisfy the scaling law and current bounds of d-type quark masses
for $\epsilon_d = \sqrt{2}$. Note that all values in the figure
are rescaled to $2$ GeV. } \label{fig:down-quark}
\end{figure}
Again it is useful to observe these values graphically. Using the
scaling law given in Eq. (\ref{eq:scaling-law}) we obtain
$m_s=\sqrt{m_b \, m_d }/\epsilon_d$. We can plot the relation as a
function of the d-quark and s-quark masses once for the highest
and once for the lowest values in $m_b(m_b)=4.25\pm 0.10$ GeV. The
output is illustrated in Fig. (\ref{fig:down-quark}). D-type quark
masses satisfying the scaling law, fall again between the two
curves. The two curves again correspond to the upper and lower
limit in the b-quark mass, which are respectively corresponding to
$m_b(m_b)=4.35$ GeV and $m_b(m_b)=4.15$ GeV and are the running
masses in the $\overline{MS}$ scheme. The grey stripe between the
dashed lines marks the strange quark mass and corresponds to the
interval $m_s(\text{1 GeV})=175\pm 0.25 $ GeV, whose highest and
lowest values are marked by the dashed lines, corresponding to
$153$  MeV and $114$ MeV at $2$ GeV respectively. The d-quark mass
$m_d=9.3 \pm 1.4$ \cite{Leut-l} is shown by the two vertical lines
which correspond to  $m_d = 6.03$ MeV to $8.17 $ MeV at $2$ GeV.

It is seen From the figure that the ranges for strange quark mass
\cite{Leut-l} and the d-quark mass \cite{Leut-l} do  have a common
intersection with the region enclosed by the curves for
$\epsilon_d=\sqrt{2}$, which is very well centered with respect to
the limits.

If we repeat the procedure by drawing  $m_s=\sqrt{m_b \, m_d
}/\epsilon_d$ with $\epsilon_d=1$  in the figure, we would have
obtained no overlap of the three regions. The constant b-quark mass curves
for $\epsilon_d=1$ lie above those plotted in the figure.

\section{The light Quark Sector}\label{sec:lqs}

It is not possible to conclude that the scaling law is consistent
with quark masses from the former two figures $alone$. One has to
make sure that  the light quark masses $m_u, m_s , m_d $
falling into the darg regions  obey $also$ the  bounds
among light quark masses obtained from current algebra. To explore
how the values of light quark masses are constrained by the
scaling law, we consider the well known relations and bounds among
light quark masses summarized in \cite{Leut-r}:
\begin{figure}[thb]
\begin{center}
\includegraphics[scale=0.8]{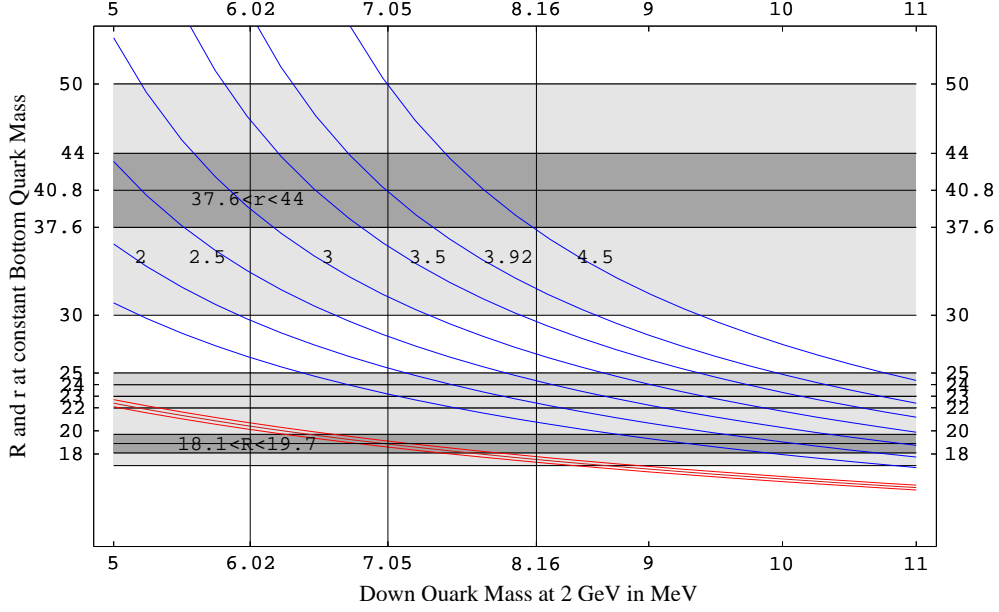}
\end{center}\caption{R and r ratios varied with respect to $m_d$ at $\epsilon=\sqrt{2}$.
The dark grey  stripes are the Leutwyler bounds and the light grey
stripes are the bounds evaluated by the particle data group.}
\label{fig:Rr}
\end{figure}
\begin{equation}\label{eq:leut}
\begin{split}
  m_u/m_d & = 0.553  \pm 0.043 \\
  m_s/m_d & = 18.9 \pm 0.8  \\
 (m_s-\overline{m}) / (m_d-m_u) & = 40.8 \pm 3.2\\
 \end{split}
\end{equation}
Note that the error bars in the values are quite small, compared
to the values evaluated by the particle data group
in~\cite{pdg-l00}. For the last two lines above we define the two
coefficients $R,r$ which are frequently used to investigate the
bounds on light quark masses
\begin{equation}\label{Rr-ratios}
\begin{split}
(i)    \ \ \ \ \  & m_s/m_d = R \\
(ii)   \ \ \ \ \  & (m_s-(m_u+m_d)/2) / (m_d-m_u) = r  \\
 \end{split}
\end{equation}
These two ratios can be graphically investigated by eliminating
$m_s$ with $\sqrt{m_b m_d}/\epsilon_d$, which follows from Eq.
(\ref{eq:scaling-law}). For the ratio R, we will use again the
upper and lower limits in $m_b$ as done before. For the ratio r,
we use only the central value of $m_b$ and since r involves the
u-quark mass, we highlight various $m_u$ values. These $m_u$
values are in turn subject to a consistency check with the scaling
done for the u-type quark masses given in Fig
(\ref{fig:up-quark}). The variation of the value of $R$ with
respect to $m_d$ and the variation of $r$ with respect to $m_d$ at
various values of $m_u$ are shown together in Fig. (\ref{fig:Rr})
where both variations are coincided with a common $m_d$ axis. This
is useful for selecting the consistent light quark masses. Note
that the variation has been done so that the masses are
constrained to exactly satisfy the scaling law.

Each of the curves in the upper half of the figure are giving the
value of r, where again each separate curve  corresponds to a
different $m_u$ value at 2 GeV. These $m_u$ values are ranging
from $1.5$ MeV to $4.5$ MeV, and are marked in the figure. Along
all of these "constant $m_u$" curves, the $central$ value of $m_b$
is also remaining constant, so that r  is a function of $m_d$ for
the specified $m_u$ and central $m_b$ mass.

The lower lying 3 adjacent curves are for $R$. Along these curves
the b-quark mass is constant such that the upper one is for the
upper limit of the b-quark mass, the lower one is for the lower
limit and the middle  one is for the central value.

\begin{figure}[thb]
\begin{center}
\includegraphics[scale=0.6]{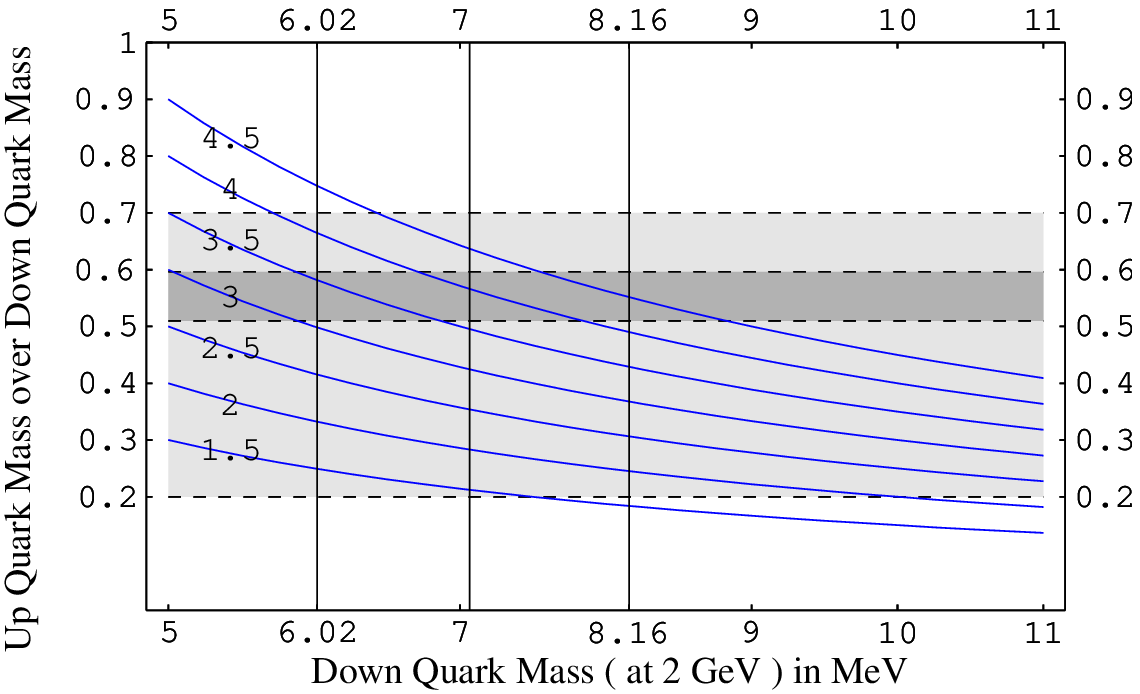}
\hspace{0.1cm} \includegraphics[scale=0.6]{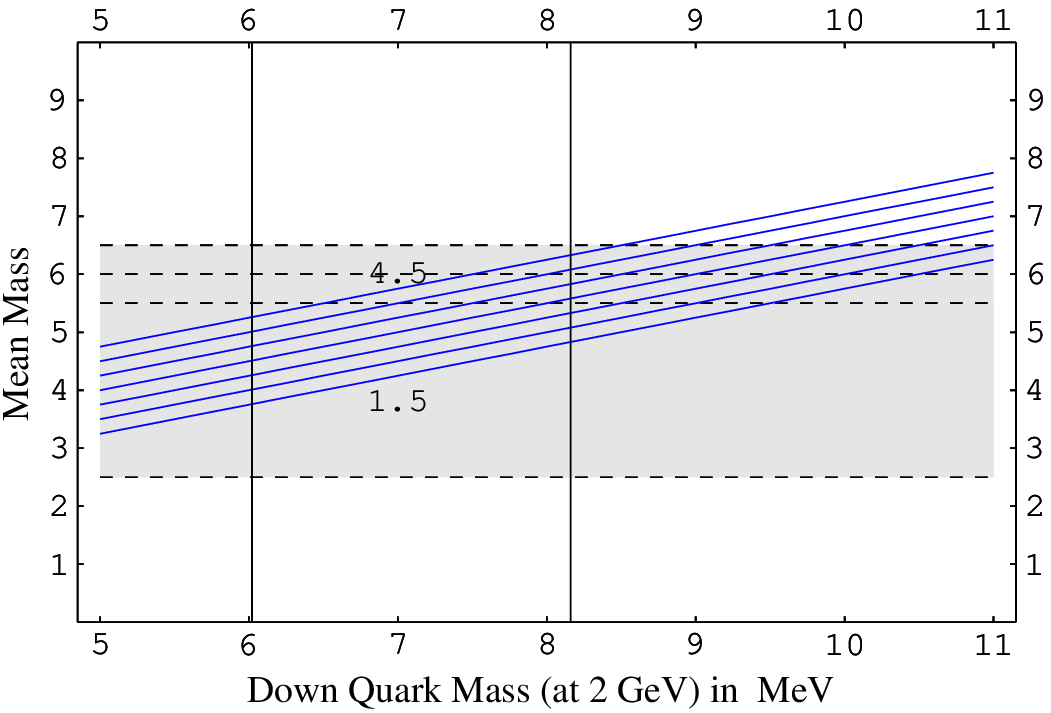}
\end{center} \caption{The ratio and the mean values of $m_u$ and $m_d$ are plotted as
a function of the mass of the down quark. The  curves correspond
to various mass values of $m_{u}$ normalized to 2 GeV. The darkest
regions is the current bound for the the ratio and the mean values
provided by \cite{Leut-r}. The lighter regions are bounds
evaluated by \cite{pdg-l00}\cite{pdg-l02}. Again for a specific
value of $m_d$ the corresponding strange quark mass and R or r
value that satisfy the scaling law can be read off from
Fig.(\ref{fig:down-quark}) and Fig.(\ref{fig:Rr}) } \label{fig:uplusd-uoverd}
\end{figure}

Before we analyze this figure, we will introduce two other
graphics which will be used in conjunction. The first shows the
variation of the ratio $m_u/m_d$ with respect to the down quark
mass $m_d$ and the second shows the variation of the mean value
$\overline{m} = (m_u+m_d) /2$ with respect to  $m_d$ , again for
various $m_u$ values that have been highlighted in fig.
(\ref{fig:Rr}). These variations are shown in Fig.
(\ref{fig:uplusd-uoverd}). The vertical lines in Fig.
(\ref{fig:uplusd-uoverd}) correspond to the highest and lowest
values in $m_d$~\cite{Leut-l} at $2$ GeV and the dark grey stripe
marks the highest and lowest  values of $m_u/m_d$ given in Eq.
(\ref{eq:leut}). In  Fig. (\ref{fig:Rr}) and
(\ref{fig:uplusd-uoverd}) the light grey regions are pdg bounds
evaluated by \cite{pdg-l00} \cite{pdg-l02} which we marked in the
figures for keeping the discussion general. Indeed it is seen that
these bounds in comparison to those given in Eq. (\ref{eq:leut})
are containing huge error bars.

All  figures in this work are produced so that they can be
combined. Indeed the behavior of the scaling law is remarkable.
Let us give an example: For a specific value of $R$ and $r$ we can
pick up a value for the d-quark mass and a value for the u-quark
mass from Fig. (\ref{fig:Rr}), then using this d-quark mass, the
corresponding strange quark mass can be read off from Fig.
(\ref{fig:down-quark}) and  the ratio $m_u/m_d$  and the mean value
$\overline{m}$ can be read from Fig. (\ref{fig:uplusd-uoverd}).
Finally the chosen $m_u$ value can be checked whether it is
consistent with the scaling of the u-type quark masses in Fig.
(\ref{fig:up-quark}).
We summarize the light and heavy quark sector for
$\epsilon_d=\sqrt{2} $ and $\epsilon_u=1$:
\begin{equation}\label{eq:lqsummary}
\begin{split}
 m_s & = 133.56  \ \ \ \     \ \    \text{MeV}   \\
 m_d & = 7.05\ \  \ \ \ \  \ \ \ \text{MeV}   \\
 m_u & = 3.92  \ \ \ \ \ \ \ \ \  \text{MeV} \\
 r   & = 40.98 \\
 R   & = 18.94 \\
 m_u/m_d & = 0.556 \\
 m_b & =  5.06 \ \ \ \ \ \ \ \ \ \text{GeV} \\
\end{split}
\end{equation}
These values  are renormalized at 2 GeV.  The bottom quark mass
rescales to $m_b(m_b)=4.28$ GeV. This perfect fitting might be an
indication that $\epsilon = \sqrt{2}$ has a physical origin in the
quark mass sector.

Using the values in eq.(\ref{eq:lqsummary}), we find  for
$m_s/\overline{m}$ and $m_s/m_u$ the values $24.34$ and $34.03$
respectively, which are also consistent with bounds
$m_s/\overline{m}=24.4 \pm 1.5 $ and  $m_s/m_u=34.4 \pm 3.7$
respectively evaluated in~\cite{Leut-r}.

Until now we used the Leutwyler and pdg bounds for quark masses as
inputs, the scaling law alone has no predictive power. Suitable
mass matrices that can give rise to such expressions might link
the scaling law with the Yukawa sector and provide a deeper
understanding. In the remaining part of the work we investigate
this possibility.

\section{A simple Model}\label{sec:model}
The set of equations in (\ref{eq:scaling-law}) are reproducible
from a simple Yukawa mass matrix which reads
\begin{equation}\label{eq:smm}
M^u =\left[\begin{array}{ccc}
  k_u & 0 & a_u \epsilon_u^\dagger  \\
  0 & a_u & 0 \\
  a_u \epsilon_u & 0 & 0 \\
\end{array}\right] , \ \ \ \ \  \ \ \ \ \
M^d =\left[\begin{array}{ccc}
  k_d & 0 & a_d  \epsilon_d^\dagger \\
  0 & a_d & 0 \\
  a_d \epsilon_d & 0 & 0 \\
\end{array}\right]
\end{equation}
where we start with the assumption that $\mid \epsilon_d \mid \,
>1$ and complex valued. Since $\epsilon_u=1$ was successful in scaling u-type quark masses
it is not necessary to impose a similar condition on $\epsilon_u$.
We also assume that $k_u,a_u$ and $k_d,a_d$  are real valued
numbers. These simple mass matrices will be later on generalized.
See also \S ~\ref{sec:permutations} for equivalent simple mass
matrices. We denote the corresponding diagonalized mass matrices
with $\mathcal{M}^u$ and $\mathcal{M}^d$. The explicit form of the
diagonal matrices are
\begin{equation}
\mathcal{M}^u=\left[\begin{array}{ccc}
  m_u & 0 & 0 \\
  0 & m_c & 0 \\
  0 & 0 & m_t \\
\end{array}\right] , \ \ \ \ \  \ \ \ \ \
 \mathcal{M}^d=\left[\begin{array}{ccc}
  m_d & 0 & 0 \\
  0 & m_s & 0 \\
  0 & 0 & m_b\\
\end{array}\right]
\end{equation}
The quark masses above are expressible through the parameters
$k_u,a_u$ and $k_d,a_d$ in the simple mass matrices as
\begin{equation}\label{eq:mass_eigenstates}
\begin{split}
 m_u & =  \frac{1}{2}\left[ k_u - \sqrt{4 \epsilon^2_u a^2_u+k^2_u}\right]  \\
 m_c & = a_u   \\
 m_t & =  \frac{1}{2}\left[ k_u + \sqrt{4 \epsilon^2_u a^2_u+k^2_u}\right]   \\
\end{split} \ \ \ \ \ \
\begin{split}
 m_d & =  \frac{1}{2}\left[ k_d  - \sqrt{4 \epsilon^2_d  a^2_d+k^2_d}\right]   \\
 m_s & = a_d    \\
 m_b & = \frac{1}{2}\left[ k_d   + \sqrt{4 \epsilon^2_d  a^2_d+k^2_d}\right]    \\
\end{split}
\end{equation}
Here  $\epsilon^2_u$ $=$ $\epsilon_u \epsilon^\dagger_u$. These
are exact expressions and  satisfy the simple scaling law
regardless of the values of the parameters.  That means, for all
values of the parameters the mass ratios are as in Eq.
(\ref{eq:scaling-law}). The parameters can be expressed in terms
of  quark masses using Eq. (\ref{eq:mass_eigenstates}). The
inverse transformations are
\begin{equation}\label{eq:ka1}
 \begin{split}
 k_u & = m_t - m_u  \  , \\
 a_u & = m_c  \  ,  \\
\end{split} \ \ \ \ \ \ \ \  \ \ \ \
 \begin{split}
 k_d & = m_b - m_d  \\
 a_d & = m_s \\
\end{split}
\end{equation}
Here it should be noted that $k_u,  k_d$ and $a_u, a_d$ are
positive. Therefore  $m_u$ and $m_d$ carry a minus sign that
follows from Eq. (\ref{eq:mass_eigenstates}). Then $k_u$ is
expressed as $m_t - m_u$ rather than $m_t + m_u$, while we assume
that masses are positive quantities. From the other side since
heavy quarks and light quarks lie in respectively GeV and MeV
scales. The $central$ values of the parameters $k_u,  k_d$ receive
some extra precision: So we have the possibility not to round the
figures in the parameters up to 6 digits, which seems to be a
useful  and  nice feature for evaluations in the CKM sector.
However the quark masses have the appropriate figures.  Note that
$k_u, k_d$ and $a_u, a_d$ are rescaled in the $\overline{MS}$
scheme, so that the ratio of the masses remain constant.

In the following part of the work we will use the simple mass
matrices to construct a satisfactory model  for quark mixing. Let
us continue with the construction of the model.

There is a transformation $V$ that diagonalizes our mass matrices
$M$ such that $\mathcal{M}=V^\dagger M V$. The diagonalizing
matrix V for the up and down simple mass matrices will be called
$V^u$ and $V^d$ respectively. They are  found as
\begin{equation}\label{eq:vuvddef}
 V^u=\left[\begin{array}{ccc}
  \cos\beta_1 & 0 & \sin\beta_1 \\
  0 & 1 & 0 \\
 -\sin\beta_1 & 0 & \cos\beta_1 \\
 \end{array}\right] \ \ \ \ \
 V^d=\left[\begin{array}{ccc}
  \cos\beta_2 & 0 & \sin\beta_2 \\
  0 & 1 & 0 \\
 -\sin\beta_2 & 0 & \cos\beta_2 \\
 \end{array}\right]
\end{equation}
where the angles $\beta_1$ and $\beta_2$ are:
\begin{subequations}
\begin{align}
  \beta_1 & = \cos^{-1} \left(-\frac{\sqrt{4 \epsilon^2_u a^2_u +k^2_u
   - k_u \sqrt{4 \epsilon^2_u a^2_u+k^2_u}}}{\sqrt{8 \epsilon^2_u a^2_u +2 \, k^2_u}
  } \right) \label{eq:beta1}\\
  \beta_2 & = \cos^{-1} \left(+\frac{\sqrt{4 \epsilon^2_d a^2_2 +k^2_d
   - k_d \sqrt{4 \epsilon^2_d a^2_2+k^2_d}}}{\sqrt{8 \epsilon^2_d a^2_d +2 \, k^2_d}
  } \label{eq:beta2} \right)
\end{align}
\end{subequations}
These are again exact expressions. The convention for the $\pm$
sign in $\beta_1$ and $\beta_2$ is discussed in
\S~\ref{sec:limit1} and \S~\ref{sec:limit2}. The resulting product
$ V^u {V^d}^\dagger$ is found as
\begin{equation}\label{eq:initial}
 V^u  \, {V^d}^\dagger = V_{\beta_1} V_{\beta_2}^\dagger = V_\delta =
\left[\begin{array}{ccc}
  \cos \delta & 0 &  \sin \delta  \\
  0 & 1 & 0 \\
  -\sin \delta  & 0 & \cos \delta  \\
\end{array}\right]
\end{equation}
Here the angle $\delta=(\beta_1-\beta_2)$. It is observed from the
entries that this matrix is not capable of reproducing the CKM
matrix~\cite{ckm} in this form. This does not imply that the
simple mass matrices giving the scaled masses are $definetly$
useless. Indeed we have the freedom to rotate the Simple Mass
Matrices\footnote{ Assume that there is a mass matrix
$\mathbb{M}^u$ that is obtained from the simple mass matrix $M^u$
in Eq. (\ref{eq:smm}) through orthogonal rotations. Then
$\mathbb{M}^u$ reduces to $M^u$ in Eq. (\ref{eq:smm}) when the
rotation specifying angles are set to some specific value as will
be made clear later. In this approach it is possible to regard the
simple mass matrices as special cases of a more general one.} in
Eq. (\ref{eq:smm}). It is seen from the $V$ matrices in Eq.
(\ref{eq:vuvddef}) that we have so far only one angle for each of
the up and down sectors\footnote{A remarkable feature of the
$\beta_1$ and $\beta_2$ angles is that for the current known
values of quark masses they appear as small deviations around
$\frac{\pi}{2}$ which is discussed in
\S~\ref{sec:pred-ckm},\S~\ref{sec:limit1} and \S~\ref{sec:limit2}
}. Indeed the generation space for each of the up and down quark
species is 3 dimensional. It tells us that each of the simple mass
matrices in Eq. (\ref{eq:smm}) must be rotated further in 2
different Euler planes , so that each of the $V$ matrices
additionally acquire two more angles, and produce an adequate
expression for the CKM matrix. Therefore we introduce further
rotations in the generation space.
    In the following it will be shown how the
mass matrices can be brought to the most general form (containing
3 angles subject to diagonalization) while keeping the mass
eigenvalues in Eq. (\ref{eq:mass_eigenstates}) $intact$. After the
Mass matrices are rotated into their final form we introduce the
complex phases as described in \S~\ref{sec:cp} and derive the
corresponding $V$ matrices that fully describe the CKM matrix. The
rotated mass matrices will be regarded as the $final$ mass
matrices of the model and the simple mass matrices in Eq.
(\ref{eq:smm}) will be considered as special cases obtainable
through setting the euler angles and complex phases to definite
values as will be shown later.

In this respect let us define the following two transformation
matrices in the generation space
\begin{equation}\label{eq:def-trans}
V_\alpha =\left[\begin{array}{ccc}
  1  & 0 & 0  \\
  0  & \cos\alpha & \sin\alpha \\
  0  & -\sin\alpha  & \cos\alpha  \\
\end{array}\right]\ \ \ \ \  \ \ \ \ \
V_\gamma =\left[\begin{array}{ccc}
  \cos\gamma  & \sin\gamma & 0  \\
  -\sin\gamma  & \cos\gamma & 0 \\
  0           & 0 & 1  \\
\end{array}\right]
\end{equation}
Note that the subscript in $V$ defines the type of rotation by
definition throughout the paper and is the argument of the sines
and cosines. Using the first one, we can rotate the mass matrix
$M^u$ into $V_{\alpha_1}^\dagger M^u V_{\alpha_1}$ and similarly
$M^d$ into $V_{\alpha_2}^\dagger M^d V_{\alpha_2}$ with arbitrary
angles $\alpha_1$ and $\alpha_2$. Now both resultant mass matrices
preserve their initial mass eigenvalues as given in
(\ref{eq:mass_eigenstates}). The transformations that diagonalize
the rotated mass matrices $V_{\alpha_1}^\dagger M^u V_{\alpha_1}$
and $V_{\alpha_2}^\dagger M^d V_{\alpha_2}$ can be reconstructed
from $V^u$ and $V^d$ through the following method
\begin{equation}\label{eq:seed}
 \begin{split}
   \mathcal{M}^u & = \left[ V_{-\alpha_1} V_{\beta_1} \right]^\dagger [V_{\alpha_1}^\dagger M^u V_{\alpha_1}] \left[ V_{-\alpha_1}
   V_{\beta_1}
   \right] \\
   \mathcal{M}^d & = \left[ V_{-\alpha_2} V_{\beta_2} \right]^\dagger [V_{\alpha_2}^\dagger M^d V_{\alpha_2}] \left[ V_{-\alpha_2}
   V_{\beta_2}
   \right] \\
\end{split}
\end{equation}
where we use the fact that $V_\alpha V_{-\alpha}$ is a unit matrix
and $V_\alpha^\dagger=V_{-\alpha}$. The modified\footnote{The term
"modified" refers to that the simple mass matrices are rotated,
and the diagonalizing transformations undergo a redefinition.}
diagonalizing transformations and modified mass matrices are then
:
\begin{equation}\label{eq:seed2}
\begin{split}
 V^u & \rightarrow [ V_{-\alpha_1} V_{\beta_1}] ,  \\
 V^d & \rightarrow [V_{-\alpha_2} V_{\beta_2}] , \\
\end{split}\ \ \ \ \ \ \
\begin{split}
 M^u & \rightarrow V_{\alpha_1}^\dagger M^u V_{\alpha_1}   \\
 M^d & \rightarrow V_{\alpha_2}^\dagger M^d V_{\alpha_2}  \\
\end{split}
\end{equation}
and the resulting modified product for quark-mixing is
\begin{equation}
\begin{split}
 V^u {V^d}^\dagger  & =  [V_{-\alpha_1}  V_{\beta_1} ] [ V_{\beta_2}^\dagger V_{-\alpha_2}^\dagger ]\\
 & = [V_{-\alpha_1} V_{\beta_1-\beta_2}  V_{\alpha_2} ] \\
 & = [V_{-\alpha_1} V_{\delta} V_{\alpha_2} ] \\
\end{split}
\end{equation}
In the last line we see that the term   is now gradually improved
with respect to  that in Eq. (\ref{eq:initial}). It contains now 4
angles. With the same token one can go a head and make use of
$V_\gamma$. Using the set of Eqs. in (\ref{eq:seed2}) we perform a
further rotation on the modified mass matrices this time applying
$V_\gamma$. Then we obtain :
\begin{equation}\label{eq:diagonalization}
 \begin{split}
   \mathcal{M}^u & = \left[ V_{-\gamma_1} V_{-\alpha_1} V_{\beta_1} \right]^\dagger \left[ V_{\gamma_1}^\dagger V_{\alpha_1}^\dagger M^u V_{\alpha_1} V_{\gamma_1}
   \right]
   \left[ V_{-\gamma_1} V_{-\alpha_1}   V_{\beta_1}  \right] \\
   \mathcal{M}^d & = \left[ V_{-\gamma_2}  V_{-\alpha_2} V_{\beta_2} \right]^\dagger \left[ V_{\gamma_2}^\dagger  V_{\alpha_2}^\dagger M^d V_{\alpha_2}
   V_{\gamma_2}\right]
   \left[ V_{-\gamma_2} V_{-\alpha_2}   V_{\beta_2}   \right] \\
\end{split}
\end{equation}
Here the modified mass matrices have  preserved their mass
eigenvalues as given in (\ref{eq:mass_eigenstates}). The
almost\footnote{The complex phase will be introduced in
\S~(\ref{sec:cp})} final transformation matrices diagonalizing
these modified mass matrices above are collectively
\begin{equation}\label{eq:no-phase}
 \begin{split}
   V^u & \rightarrow  V_{-\gamma_1} V_{-\alpha_1}   V_{\beta_1}  \ , \\
   V^d & \rightarrow  V_{-\gamma_2} V_{-\alpha_2}   V_{\beta_2} \  , \\
 \end{split} \   \ \ \ \ \ \ \ \
 \begin{split}
   M^u & \rightarrow   V_{\gamma_1}^\dagger V_{\alpha_1}^\dagger M^u V_{\alpha_1} V_{\gamma_1}   \\
   M^d & \rightarrow   V_{\gamma_2}^\dagger V_{\alpha_2}^\dagger M^d V_{\alpha_2}  V_{\gamma_2} \\
\end{split}
\end{equation}
respectively. From Eq. (\ref{eq:no-phase}), the almost final form
of the CKM matrix can be written as
\begin{equation}
 \begin{split}
   ( V^u ) \, (V^d)^\dagger & =  \left[ V_{-\gamma_1} V_{-\alpha_1}   V_{\beta_1}  \right]
                           \left[V_{-\gamma_2} V_{-\alpha_2}   V_{\beta_2}  \right]^\dagger \\
                     & =   V_{-\gamma_1} V_{-\alpha_1}   V_{\beta_1}
   V_{\beta_2}^\dagger  V_{-\alpha_2}^\dagger  V_{-\gamma_2}^\dagger    \\
                        & =   V_{-\gamma_1} V_{-\alpha_1}   V_{\delta}
     V^\dagger _{-\alpha_2} V^\dagger_{-\gamma_2}    \\
                        & =   V_{-\gamma_1} V_{-\alpha_1}   V_{\delta}
     V_{\alpha_2} V_{\gamma_2}    \\
\end{split}
\end{equation}
It is seen at first sight that the six angles that meet each other
in the expression induce an asymmetry. So we achieved the
mentioned point. There are 6 angles in the CKM matrix and the
rotated mass matrices generate the scaling law expression with the
mass eigenvalues in Eq.~(\ref{eq:mass_eigenstates}). The simple
mass matrices in Eq.~(\ref{eq:smm}) can be obtained by setting
$\alpha_1=\alpha_2=\pi$ and $\gamma_1=\gamma_2=\pi$ in
Eq.~(\ref{eq:no-phase}).

To enable a comparison with the standardized CKM matrix, we just
let $\delta$ temporarily be zero, then $V_{\delta}$ becomes a unit
matrix and we obtain
\begin{equation*}
\left[\begin{array}{ccc}
  \cos\gamma_1  & -\sin\gamma_1 & 0  \\
  \sin\gamma_1  & \cos\gamma_1 & 0 \\
  0           & 0 & 1  \\
\end{array}\right]
\left[\begin{array}{ccc}
  1  & 0 & 0  \\
  0  & \cos \epsilon & -\sin\epsilon \\
  0  & \sin\epsilon  & \cos\epsilon  \\
\end{array}\right]
\left[\begin{array}{ccc}
   \cos\gamma_2  & \sin\gamma_2 & 0  \\
  -\sin\gamma_2  & \cos\gamma_2 & 0 \\
  0           & 0 & 1  \\
\end{array}\right]
\end{equation*}
where  we have rewritten the term $ V_{-\alpha_1} V_{\alpha_2}$ as
$V_{\epsilon} $ such that $\epsilon = (\alpha_1-\alpha_2) $. The
above form is equivalent to the standard form of the $CKM$ matrix
given in \cite{Fritzsch}. If we consider the PDG version of the
$CKM$ matrix~\cite{pdg-ckm} we see that it has 3 angles and a
phase. We defined totally $2 \, \alpha $'s,  $ 2 \, \gamma$'s and
$2 \, \beta $'s which makes totally 6 angles. It should be noted
that this is not an over parametrization. Indeed the relative
values count, this makes than 3 parameters. The extra phases that
give rise to $CP$ violation will be introduced in a similar
fashion in the next section. Here of course nothing prohibits us
from fixing the 6 angles $(-\gamma_1
,-\alpha_1,\beta_1,-\beta_2,\alpha_2,\gamma_2 )$ to the
experimental values. But  the parameter $\delta$ is related with
the quark masses\footnote{Note that $\delta$ is the only angle
that determines the quark masses, since $\beta_1$ and $\beta_2$
are functions of $k_u,a_u$ and $k_d,a_d$ respectively. Other
angles are due to rotations in the generation space. It is
discussed in \S~(\ref{sec:permutations}) how other choices of
simple mass matrices are possible that give equivalent
descriptions of the CKM matrix. In such equivalent descriptions
$\delta$ will be replaced by rotation angles operating in other
rotation planes.} and is not a completely free parameter. It is of
interest whether $\delta$ will consistently predict the $CKM$
entries. In order  to see the effect of $\delta$ on the entries,
in the following part of the work we expand the expression in a
series, for small angles. The expanded form is then adequate for a
comparison with the Wolfenstein parametrization
\cite{Wolfenstein}. Let us first introduce the $CP$ violating
phase in our model.

\section{CP violating phase }\label{sec:cp}
During the construction of the $CKM$ matrix we had ignored the
quark phases. A suitable way to incorporate the phases is to
modify the mass matrices. We use the equations in
(\ref{eq:diagonalization}) and introduce the quark phases
\begin{equation}\label{eq:mass-matrices}
 \begin{split}
   \mathcal{M}^u & = \left[V_{-\gamma_1}  V_{-\alpha_1} V_{\phi_1} V_{\beta_1} \right]^\dagger \left[  V_{\gamma_1}^\dagger V_{\alpha_1}^\dagger V_{\phi_1}  M^u V_{\phi_1}^\dagger V_{\alpha_1} V_{\gamma_1}
   \right]
   \left[ V_{-\gamma_1} V_{-\alpha_1}  V_{\phi_1} V_{\beta_1}  \right] \\
   \mathcal{M}^d & = \left[ V_{-\gamma_2}  V_{-\alpha_2} V_{\phi_2} V_{\beta_2} \right]^\dagger \left[ V_{\gamma_2}^\dagger  V_{\alpha_2}^\dagger V_{\phi_2} M^d V_{\phi_2}^\dagger V_{\alpha_2}
   V_{\gamma_2}\right]
   \left[ V_{-\gamma_2} V_{-\alpha_2} V_{\phi_2}  V_{\beta_2}   \right] \\
\end{split}
\end{equation}
The middle terms in the brackets $[...]$ are the $final$ mass
Matrices for up and down quarks.  A suitable choice for the
transformations $V_{\phi_2}$ and $V_{\phi_1}$ might be simply a
diagonal matrix with quark phases as entries
\begin{equation}\label{eq:phases}
V_{\phi_1} =\left[\begin{array}{ccc}
  e^{i \, \phi_u }  & 0 & 0 \\
  0  & e^{i \, \phi_c}  & 0 \\
  0  & 0 & e^{i \, \phi_t}  \\
\end{array}\right]\ \ \ \ \  \ \ \ \ \
V_{\phi_2} =\left[\begin{array}{ccc}
  e^{i \, \phi_d }  & 0 & 0 \\
  0  & e^{i \, \phi_s}  & 0 \\
  0  & 0 & e^{i \, \phi_b}  \\
\end{array}\right]
\end{equation}
The expressions for the diagonalizing transformations $V^u$ and
$V^d$ are now containing the phase information as well. The final
form of the $V$ matrices and mass matrices are :
\begin{equation}\label{eq:f}
 \begin{split}
     V^u & \rightarrow  V_{-\gamma_1} V_{-\alpha_1}  V_{\phi_1} V_{\beta_1}  , \\
     V^d & \rightarrow  V_{-\gamma_2} V_{-\alpha_2}  V_{\phi_2}  V_{\beta_2}  , \\
\end{split} \   \ \ \ \ \ \ \ \
 \begin{split}
   M^u & \rightarrow   V_{\gamma_1}^\dagger V_{\alpha_1}^\dagger V_{\phi_1}  M^u V_{\phi_1}^\dagger V_{\alpha_1} V_{\gamma_1}
    \\
   M^d & \rightarrow  V_{\gamma_2}^\dagger  V_{\alpha_2}^\dagger V_{\phi_2} M^d V_{\phi_2}^\dagger V_{\alpha_2}
   V_{\gamma_2}  \\
\end{split}
\end{equation}
The $CKM$ matrix in our model takes its final form as
\begin{equation}\label{eq:exact-ckm}
 \begin{split}
   ( V^u ) \, (V^d)^\dagger  & = U^{CKM}= V_{-\gamma_1} V_{-\alpha_1} V_{\phi_1}
   V_{\delta} {V_{\phi_2}}^\dagger
     V_{\alpha_2} V_{\gamma_2}   \\
\end{split}
\end{equation}
Note that when we temporarily set $\delta$ and $ \gamma_2 $ to
zero and collect $V_{\phi_1}$ and $V^\dagger_{\phi_2}$ in a single
expression with one non-vanishing phase such that $V_{\phi_1}
V^\dagger_{\phi_2} = diag[1,1,e^{i \,( \phi_t-\phi_b )}]$, then we
obtain a rather standard form.
\begin{equation*}
\begin{small}
\left[\begin{array}{ccc}
  c_{\gamma_1}  & -s_{\gamma_1} & 0  \\
  s_{\gamma_1}  & c_{\gamma_1} & 0 \\
  0           & 0 & 1  \\
\end{array}\right]
\left[\begin{array}{ccc}
  1  & 0 & 0  \\
  0  & c_{\alpha_1} & -s_{\alpha_1} \\
  0  & s_{\alpha_1}  & c_{\alpha_1} \\
\end{array}\right]
\left[\begin{array}{ccc}
  1  & 0  & 0  \\
  0  & 1  & 0 \\
  0  & 0  & e^{i\phi_{tb}} \\
\end{array}\right]
\left[\begin{array}{ccc}
    c_{\gamma_2}  & s_{\gamma_2} & 0  \\
   -s_{\gamma_2}  &  c_{\gamma_2} & 0 \\
  0           & 0 & 1  \\
\end{array}\right]
\end{small}
\end{equation*}
where we shortly denote $\phi_{tb}=\phi_t-\phi_b$. This shows how
close our model stands to the standardized form. The main
difference stems from $\delta$ as discussed before, which is
determinable from quark masses and the phases $\phi_u , \phi_c ,
\phi_t ,$ and $\phi_d , \phi_s   , \phi_b $ which follow from
observable $CP$ violation.

\section{Predicting the CKM entries}\label{sec:pred-ckm}

At the first stage we are interested in what influence
$V_{\delta}$ might have. We use the quark masses summarized in \S
(\ref{sec:sqm-dtype}) , \S(\ref{sec:sqm-utype}) and
\S(\ref{sec:lqs}) to determine the parameters $k_u , a_u$ and $k_d
, a_d$ given in Eq.(\ref{eq:ka1}) and subsequently insert them
into Eq.(\ref{eq:beta1}) and into Eq. (\ref{eq:beta2}) to obtain
$\beta_1$ and $\beta_2$. It is remarkable that these values appear
as small deviations around $\frac{\pi}{2}$. This is discussed in
\S \ref{sec:limit1} and \S \ref{sec:limit2}. As $\beta_1$ and
$\beta_2$ are small, $\delta$ will also be small. We obtain
$\delta = \beta_1 -\beta_2 = 0.040868$. Here the precision comes
from the unrounded figures in $k_u,k_d$, which was discussed
before. Let us start with
\begin{equation}\label{eq:ckm-th}
 \begin{split}
   ( V^u ) \, (V^d)^\dagger  & = U^{CKM}= V_{-\gamma_1} V_{-\alpha_1}   V_{\phi_1}
   V_{\delta} {V_{\phi_2}}^\dagger
     V_{\alpha_2} V_{\gamma_2}   \\
\end{split}
\end{equation}
For explicit calculations, the matrices $V_{-\gamma_1}$ , $
V_{-\alpha_1}$ ,$V_{\alpha_2}$ and  $V_{\gamma_2}$  are defined as
in Eq. (\ref{eq:def-trans}) and $V_{\delta}$ is given in  Eq.
(\ref{eq:initial}).  With some trial\footnote{Note that the model
does not predict the angles. We choose those  values for the
angles which reproduce the current values of the CKM matrix. It
seems at first stage that the model is trivial, but that is not
the case since once $\delta$ is fixed to the known masses of
quarks the remaining angles do not present sufficient freedom to
fix any arbitrary CKM matrix. On the other side the values of
$k_u,a_u$ and $k_d, a_d$ which determine $\delta$ have a physical
origin as discussed in \S~\ref{sec:vevs} and are related with a
mass generating mechanism.} we choose for the parameters,
\begin{eqnarray}\label{eq:central-values}
 \begin{array}{lcrcr}
   \gamma_1 & = & \frac{\pi}{2} + 0.226580   & = & \frac{\pi}{2} + \Delta_{\gamma_1} \\
   \gamma_2 & = & \frac{\pi}{2} + 0.003000   & = & \frac{\pi}{2} + \Delta_{\gamma_2} \\
   \alpha_1 & = &  0   +  0.030000           & = & 0+  \Delta_{\alpha_1}\\
   \alpha_2 & = &  0   +  0.022658           & = & 0+  \Delta_{\alpha_2}\\
   \beta_1  & = & \frac{\pi}{2} +0.003557    & = & \frac{\pi}{2}-\Delta_{\beta_1}\\
   \beta_2  & = & \frac{\pi}{2} -0.037312    & = & \frac{\pi}{2}-\Delta_{\beta_2}\\
\end{array}
\end{eqnarray}
Where the $\Delta$ 's denote the amount of deviation from the
central values.  We simply substitute these values in eq.(\ref{eq:ckm-th}).
The absolute values are found as
\begin{equation}\label{eq:ckm}
    \left[\begin{array}{lll}
     0.974867    & 0.222759    & 0.003651 \\
     0.222640    & 0.974017    & 0.041505 \\
     0.008155    & 0.040859    & 0.999132 \\
    \end{array}\right]
\end{equation}
Where a complex phase $\phi_c = 0.144$ has been used, while all
other phases are set identically to zero. The above values of the
parameters reproduce the $CKM$ entries in an excellent agreement
with the current values~\cite{pdg-ckm}.
\begin{figure}[t]
\begin{center}
\includegraphics[scale=0.5]{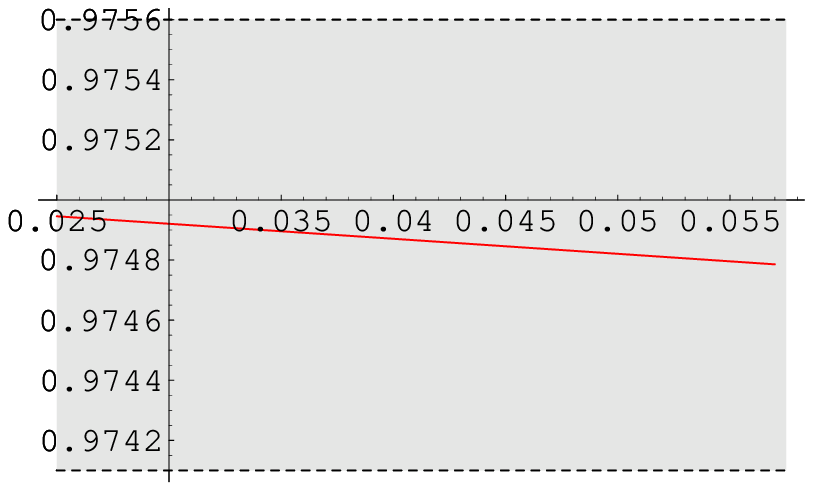}
\includegraphics[scale=0.5]{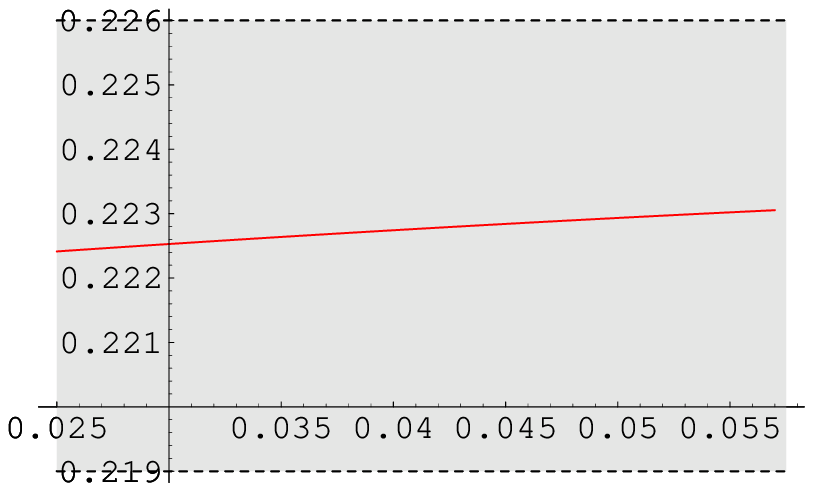}
\includegraphics[scale=0.5]{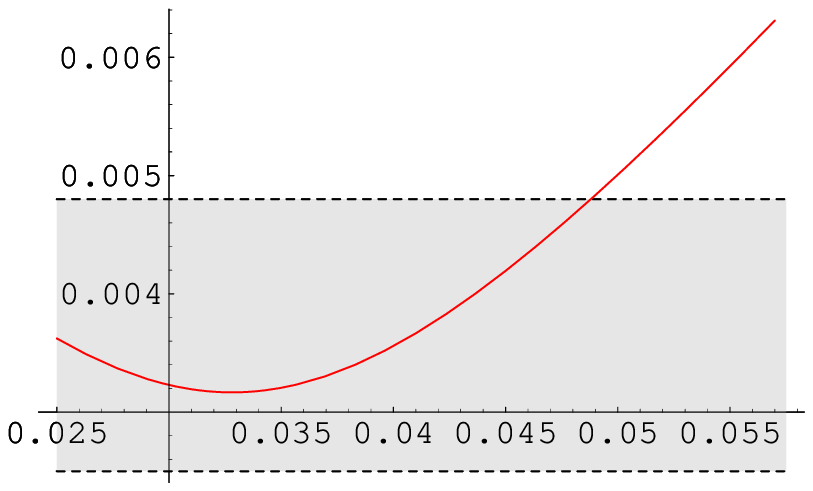}
\end{center}
\begin{center}
\includegraphics[scale=0.5]{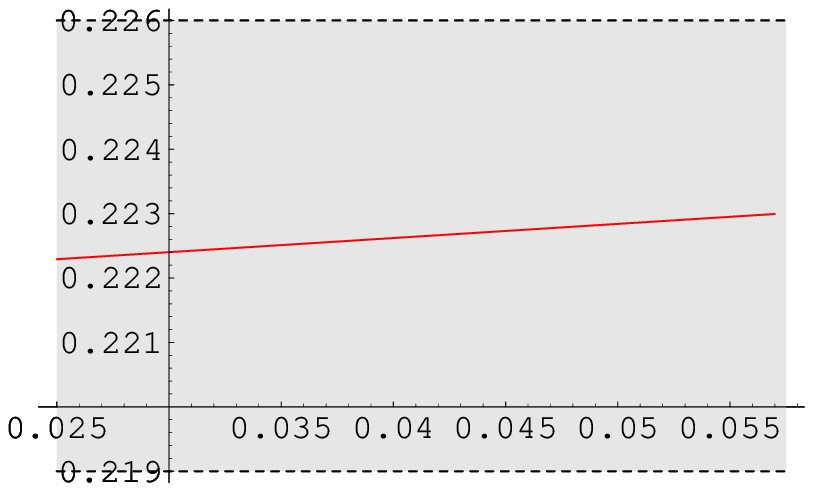}
\includegraphics[scale=0.5]{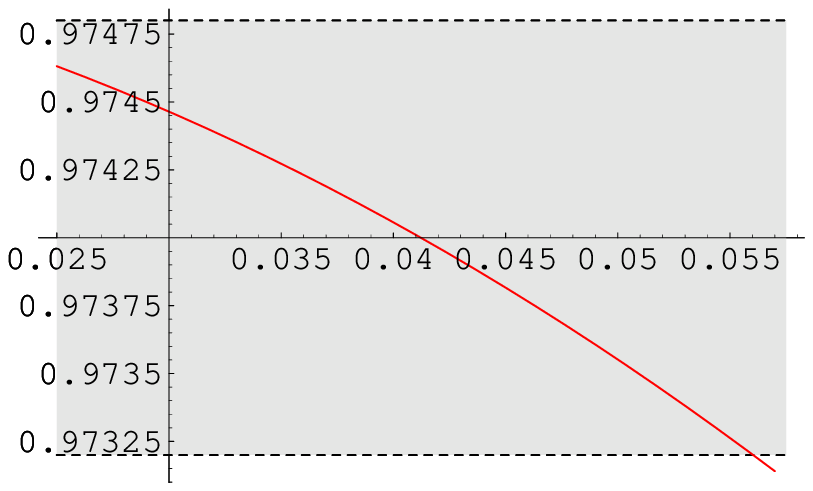}
\includegraphics[scale=0.5]{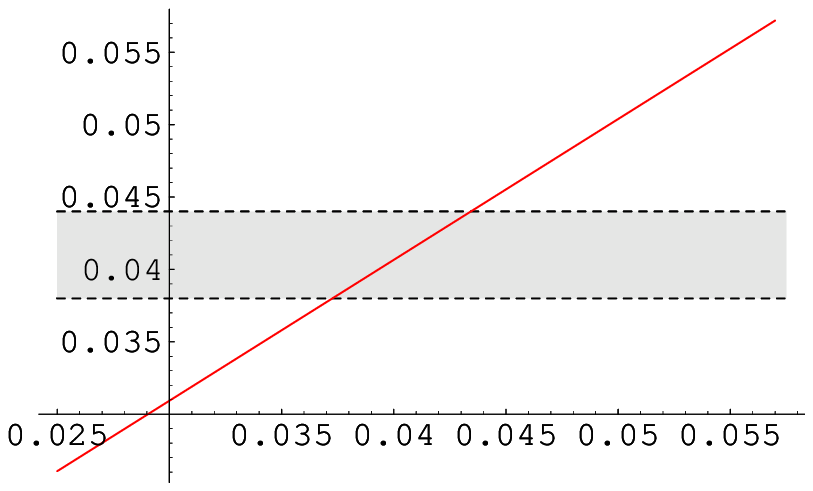}
\end{center}
\begin{center}
\includegraphics[scale=0.5]{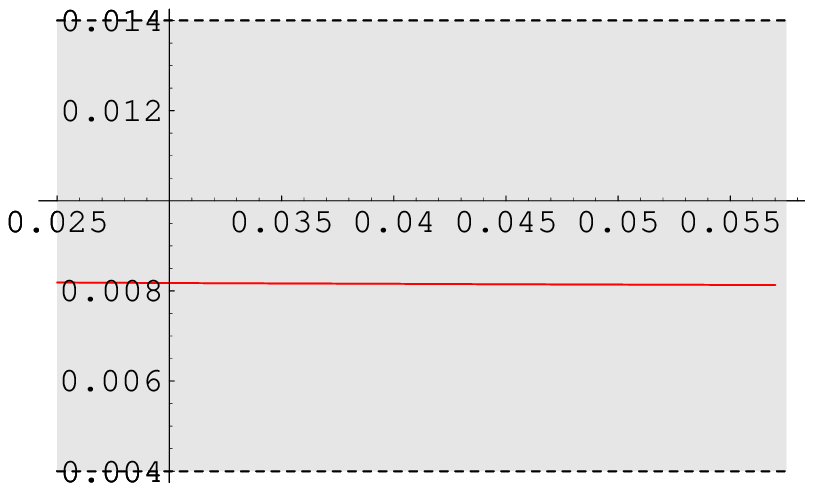}
\includegraphics[scale=0.5]{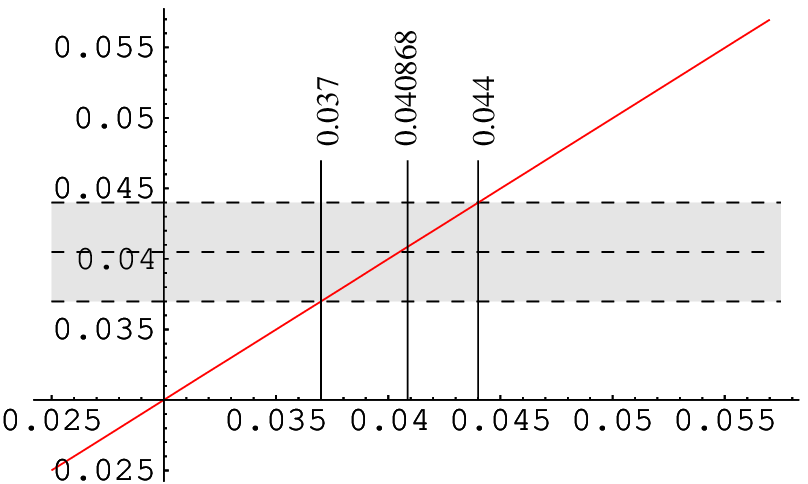}
\includegraphics[scale=0.5]{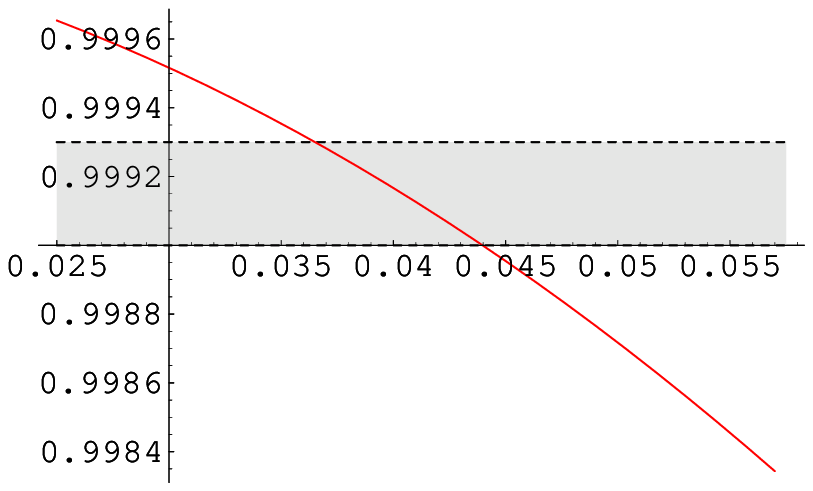}
\end{center} \caption{Variation of $CKM$ entries with respect to  $\delta$, the position
of each figure overlaps with  its position in the $CKM$
matrix.}\label{fig:ckm-graph}
\end{figure}
Slight changes are possible through changing the central values of
the angles at fixed $\delta = 0.040868$. Whether the model
contains any triviality with respect to the parameter $\delta$
could be clarified by investigating the effect of $\delta$ on the
$CKM$ entries. For example would it be  possible to fix the angles
$(-\gamma_1 ,-\alpha_1,\alpha_2,\gamma_2 )$ so that they give  the
current $CKM$ values regardless of what value $\delta$ takes ?

\begin{figure}[thb]
\begin{center}
\includegraphics[scale=0.5]{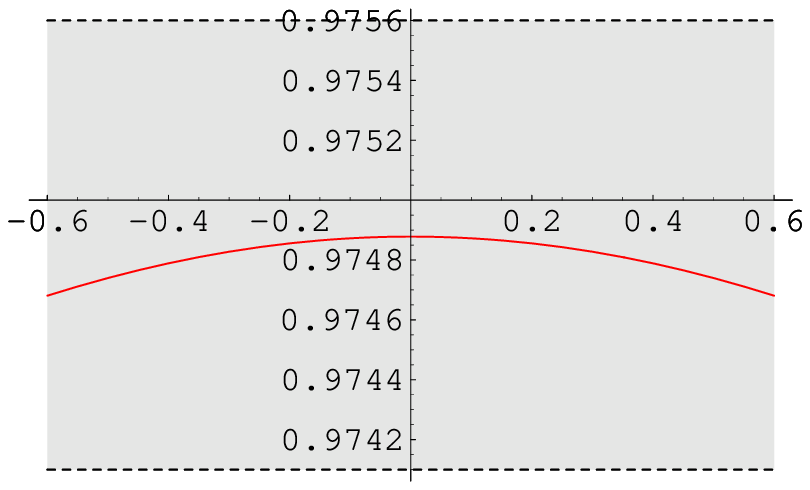}
\includegraphics[scale=0.5]{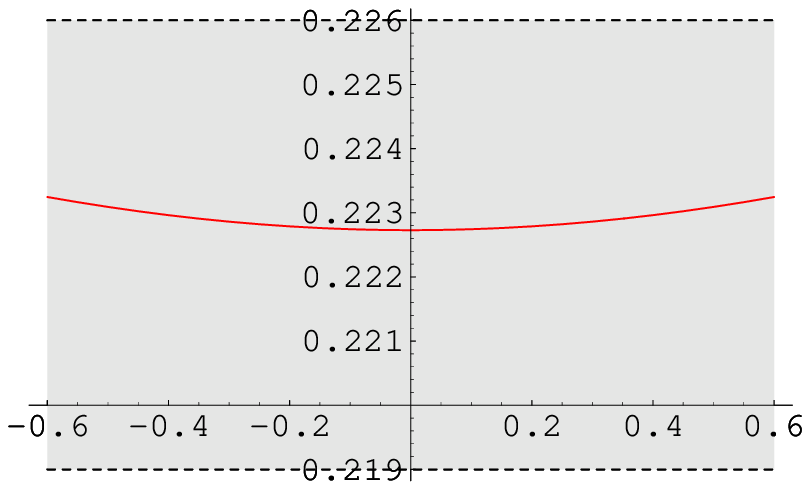}
\includegraphics[scale=0.5]{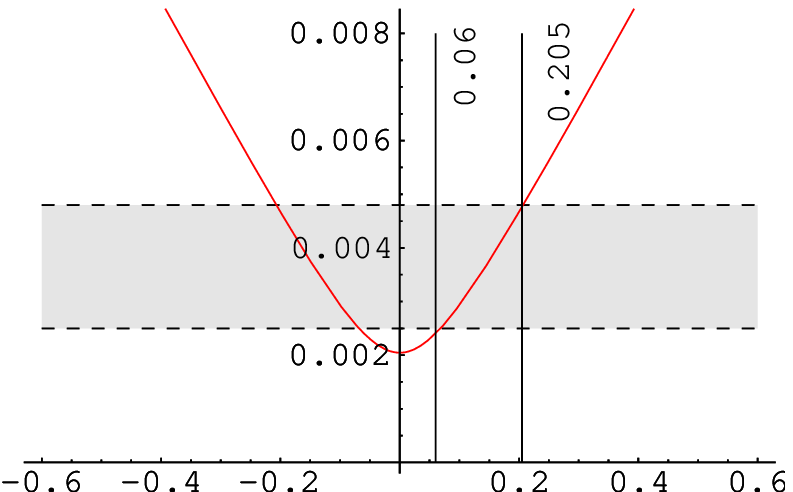}
\end{center}
\begin{center}
\includegraphics[scale=0.5]{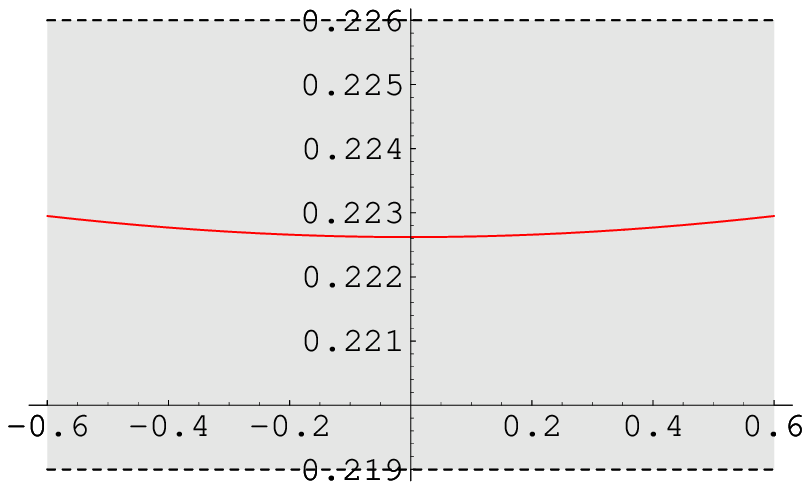}
\includegraphics[scale=0.5]{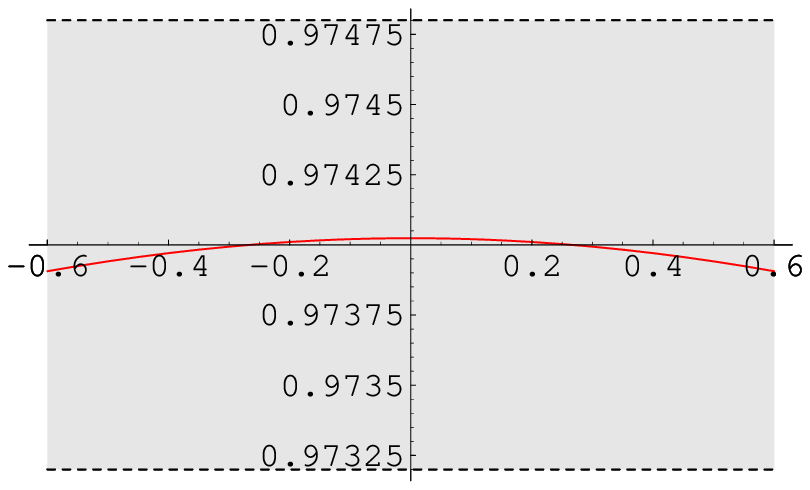}
\includegraphics[scale=0.5]{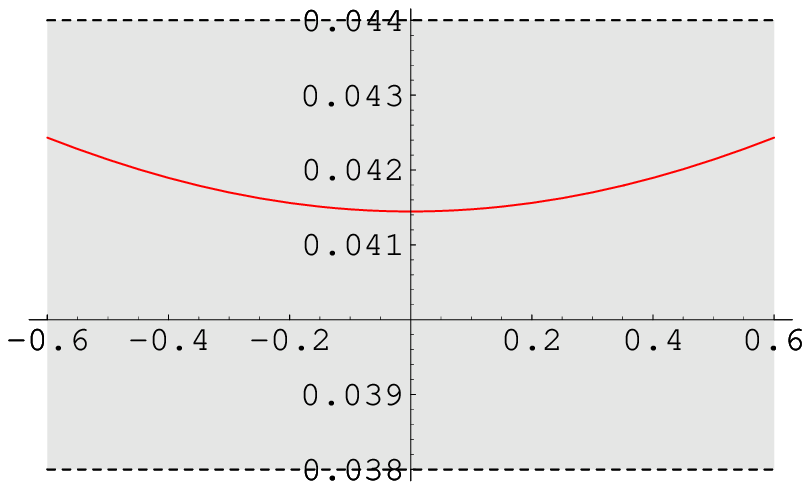}
\end{center}
\begin{center}
\includegraphics[scale=0.5]{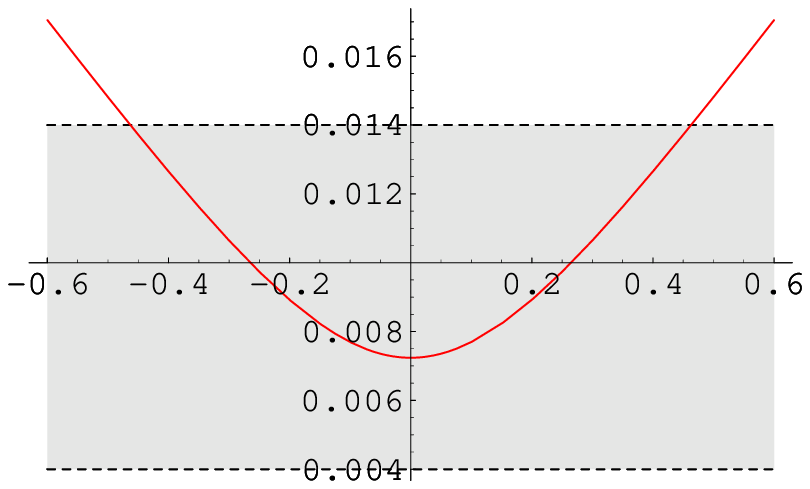}
\includegraphics[scale=0.5]{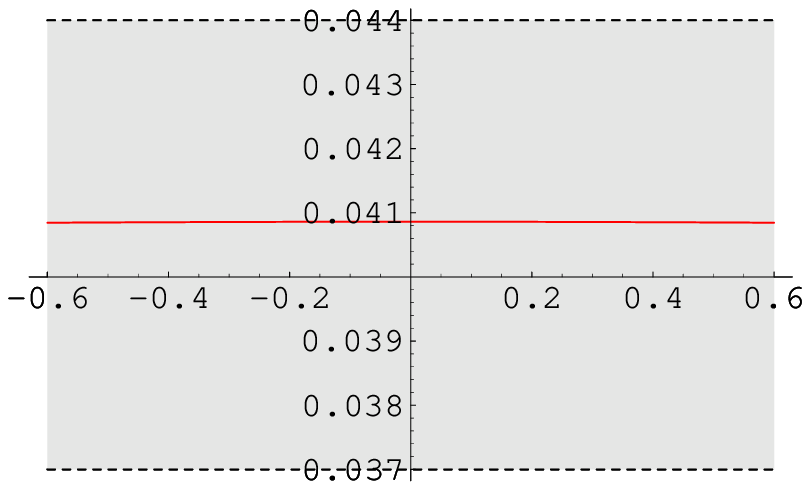}
\includegraphics[scale=0.5]{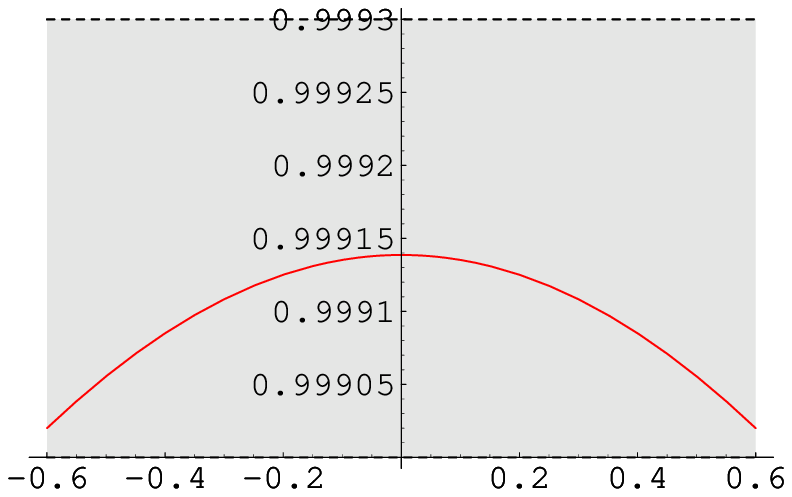}
\end{center} \caption{Variation of the $CKM$ entries with respect to  $\phi_c$, the position
of each figure overlaps with  its position in the $CKM$
matrix.}\label{fig:ckm-graph-phase-c}
\end{figure}

A graphical approach shows that the central values of the  $CKM$
entries are reproducible only within a very narrow band for
$\delta$ which lies approximately in the range $0.037<\delta
<0.044$. This fact is illustrated in  fig. (\ref{fig:ckm-graph})
where all other parameters are kept fixed as in Eq.
(\ref{eq:central-values}), but only $\delta$ is varied over   a
large interval, starting from $0.025$ up to $0.057$. Indeed this
interval is too large and  will  produce bad quark masses as we
depart from the central value:  $\delta = 0.040868$. As seen from
the figures, the entries $V_{ts}$ ,$V_{cb}$ , $V_{tb}$ are largely
$\delta$ dependent. The grey regions are current bounds for the
entries. Again from the variation of the CKM entries with respect
to  $\delta$, we see that the bound on $\delta$ is largely imposed
by the $V_{ts}$ ,$V_{cb}$ , $V_{tb}$ entries. The best values are
in the interval $0.037<\delta <0.044$. This means that the better
these entries are known the more feedback is obtained for
determining quark masses. Our previous  analysis for quark masses
gave a rather good value.i.e.,  $\delta = 0.040868$ which is both
consistent with the $CKM$ entries and quark masses. It also allows
a total phase of $\phi_c=0.144$ which is consistent. We consider
as next the variation of the $CKM$ entries with respect to the
phases $(\phi_u , \phi_c , \phi_t )$ at constant values of the parameters as given
in  Eq.
(\ref{eq:central-values}).
\paragraph{\bf Case A :}
The variation of $CKM$ entries with respect to  $\phi_c$ in  the
large interval  $-0.6<\phi_c<0.6$ are shown in Fig.
(\ref{fig:ckm-graph-phase-c}). Here all other phases are set
identically to zero.
It is seen that $V_{ub}$ and $V_{td}$ are largely $\phi_c$
dependent. The determination of the $CP$ violating phase is
therefore predictable from precise measurements of these entries.
The current range for $V_{ub}$ which is $0.0025\dots0.0048$
constrains $\phi_c$ to vary between $0.06<\phi_c<0.205$ as seen
from the figure.
\paragraph{\bf Case B :} Let us set all phases to zero
and  vary this time $\phi_u$ in the interval $-0.6<\phi_u<06$.
Again the current value of $V_{ub}$  constrains $\phi_u$ to vary
between $0.177<\phi_u<0.543$ as seen from the figure.
\begin{figure}[thb]
\begin{center}
\includegraphics[scale=0.5]{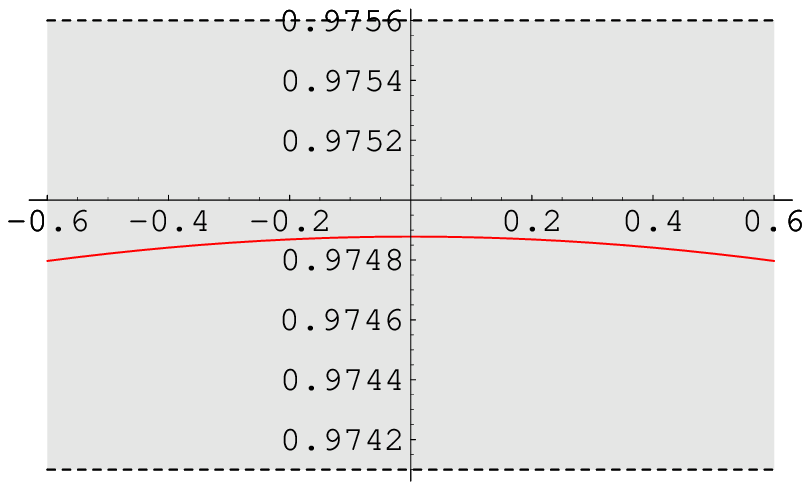}
\includegraphics[scale=0.5]{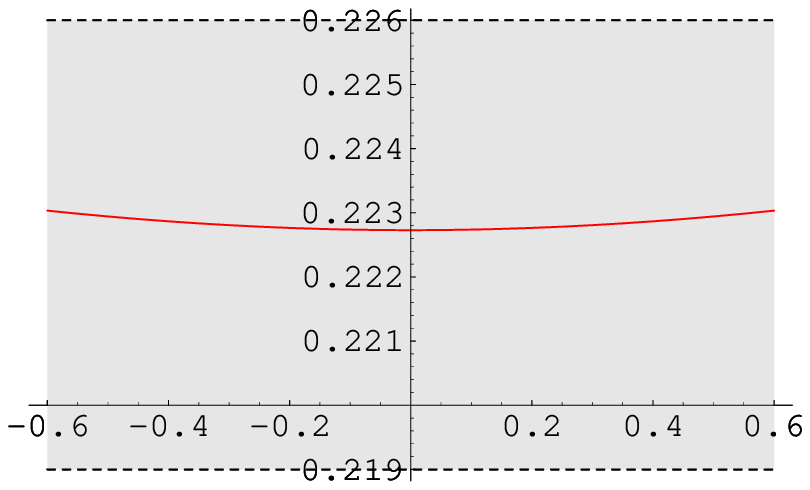}
\includegraphics[scale=0.5]{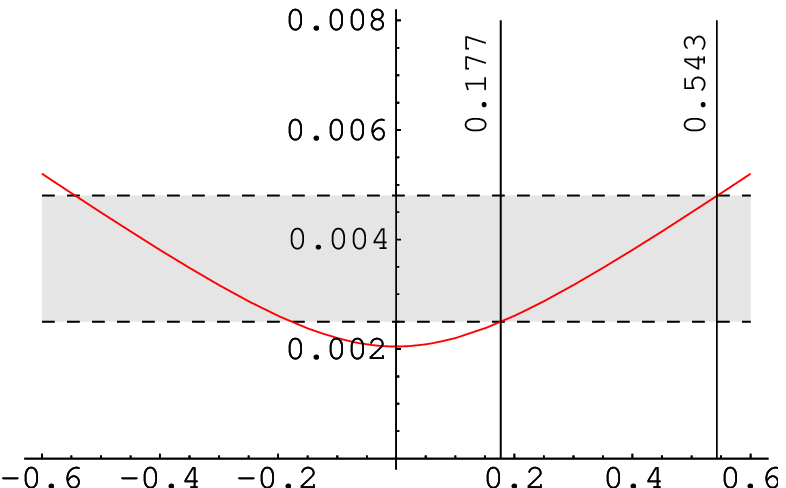}
\end{center}
\begin{center}
\includegraphics[scale=0.5]{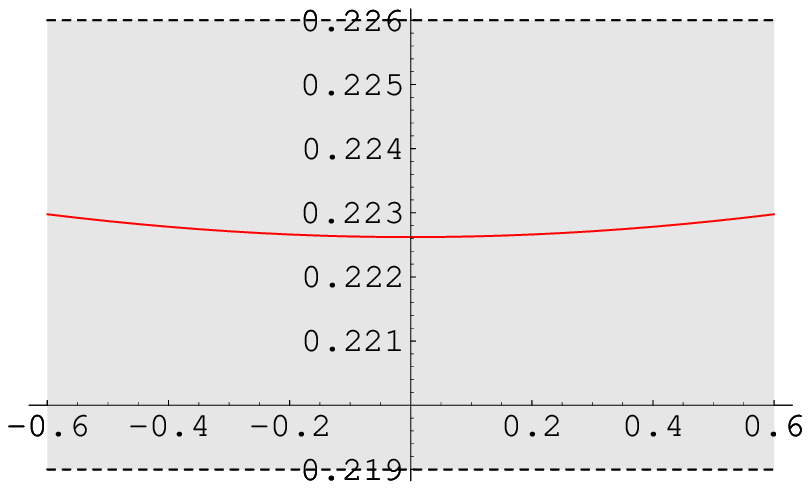}
\includegraphics[scale=0.5]{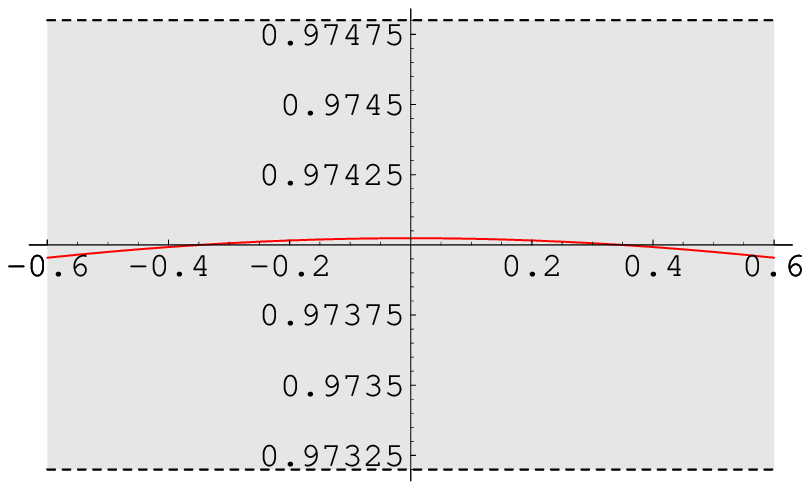}
\includegraphics[scale=0.5]{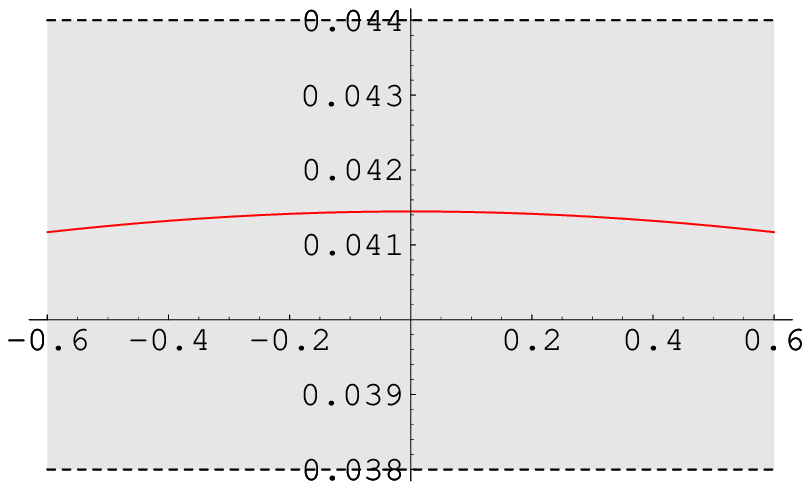}
\end{center}
\begin{center}
\includegraphics[scale=0.5]{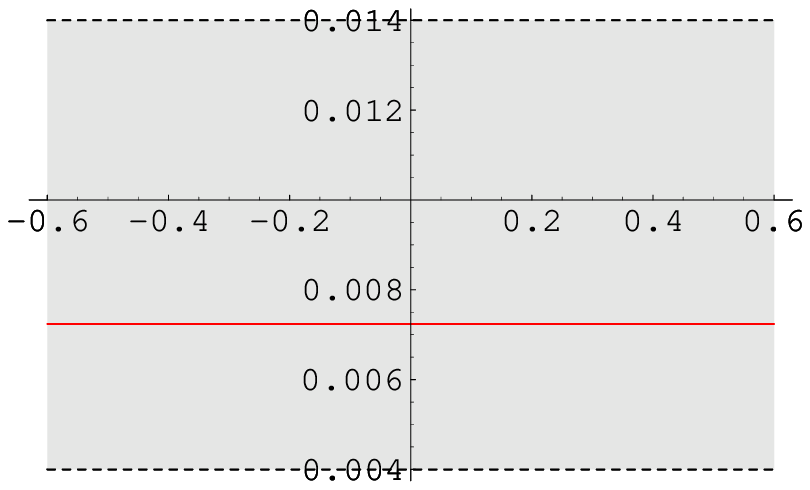}
\includegraphics[scale=0.5]{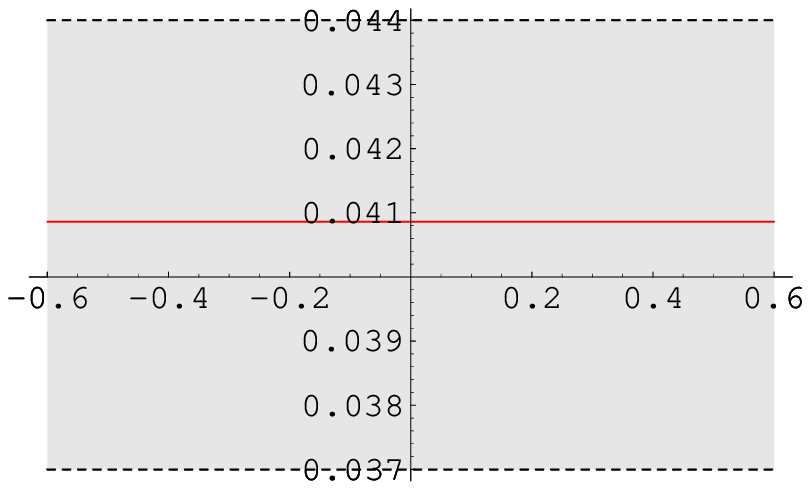}
\includegraphics[scale=0.5]{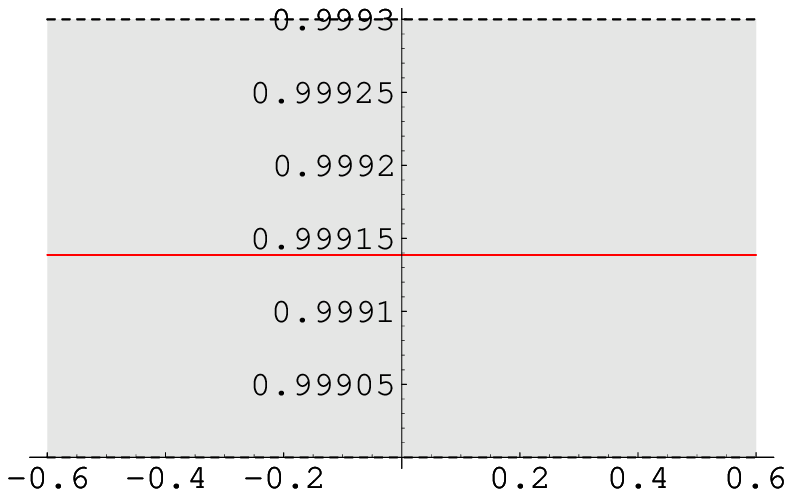}
\end{center} \caption{Variation of $CKM$ entries w.r.t  the parameter $\phi_u$, the position
of each figure overlaps with  its position in the $CKM$
matrix.}\label{fig:ckm-graph-phase-u}
\end{figure}
\paragraph{\bf Case C :}
A final case that we illustrate is where only $\phi_t$ is varied.
Again all other phases  are set to zero and we vary $\phi_t$ in the
range $-0.6<\phi_t<0.6$. This time the current value of $V_{ub}$
severely constrains $\phi_t$ to the interval $0.046<\phi_t<0.142$ as seen from
the figure. \\
\begin{figure}[thb]
\begin{center}
\includegraphics[scale=0.5]{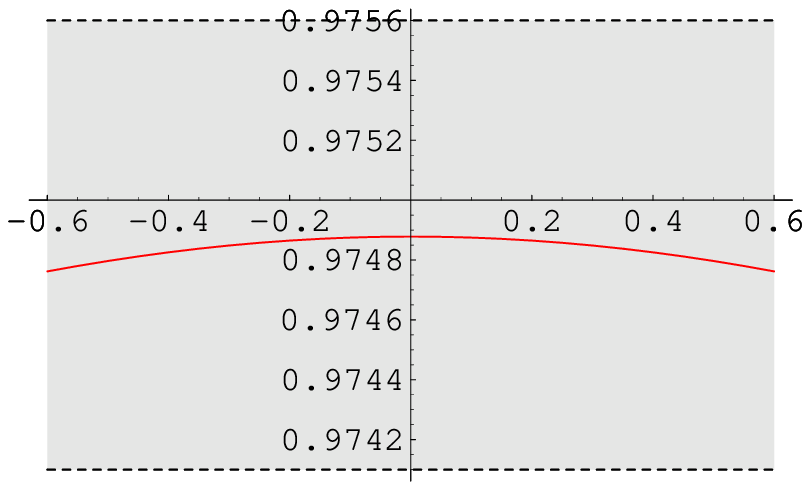}
\includegraphics[scale=0.5]{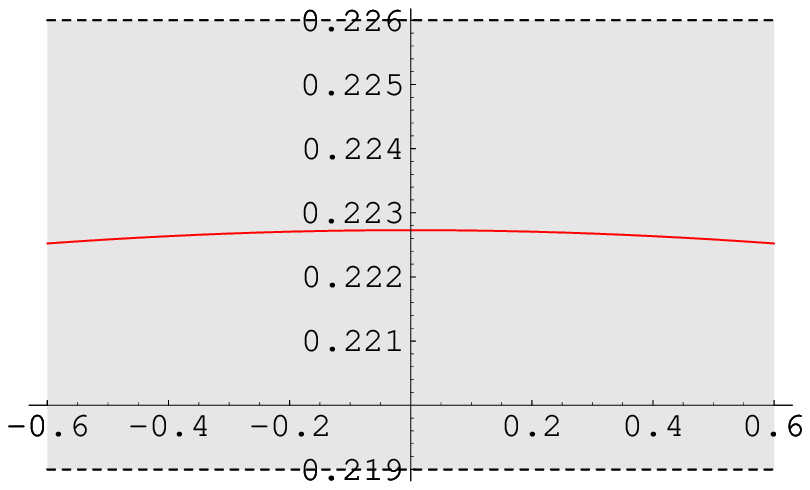}
\includegraphics[scale=0.5]{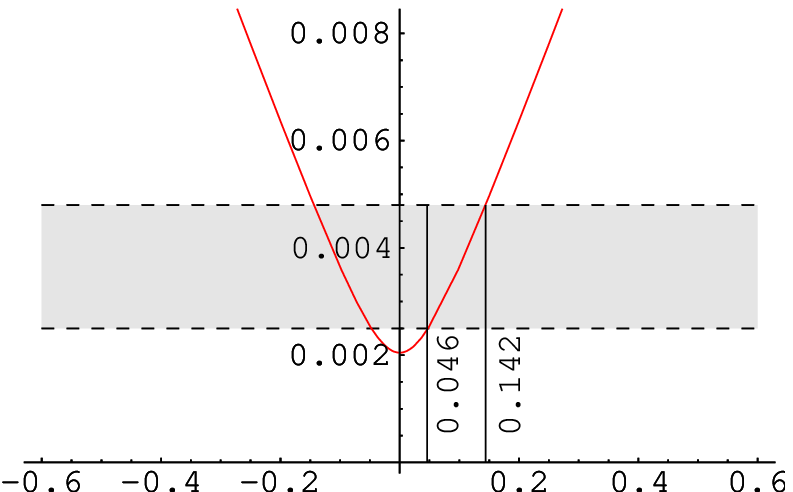}
\end{center}
\begin{center}
\includegraphics[scale=0.5]{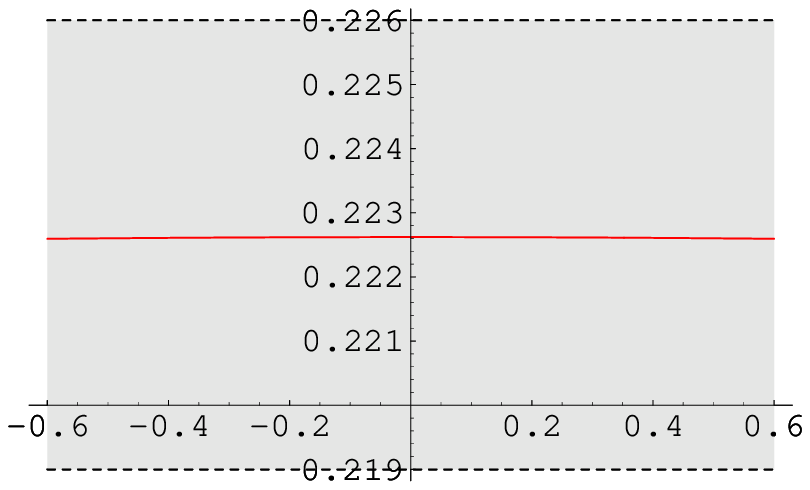}
\includegraphics[scale=0.5]{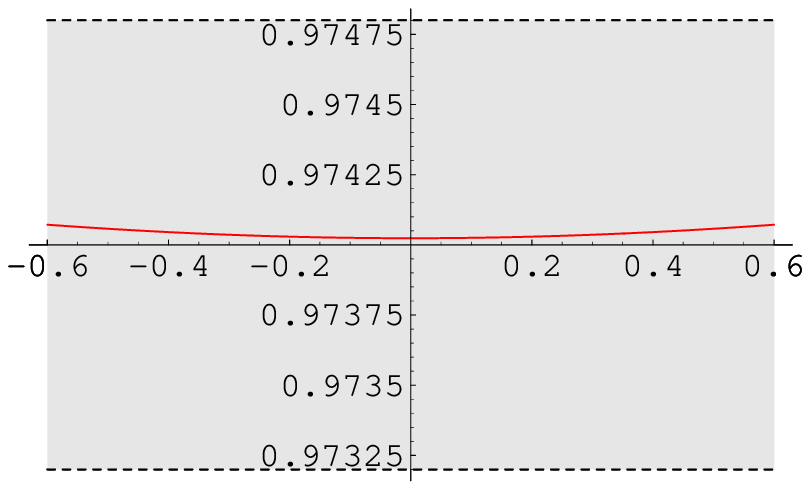}
\includegraphics[scale=0.5]{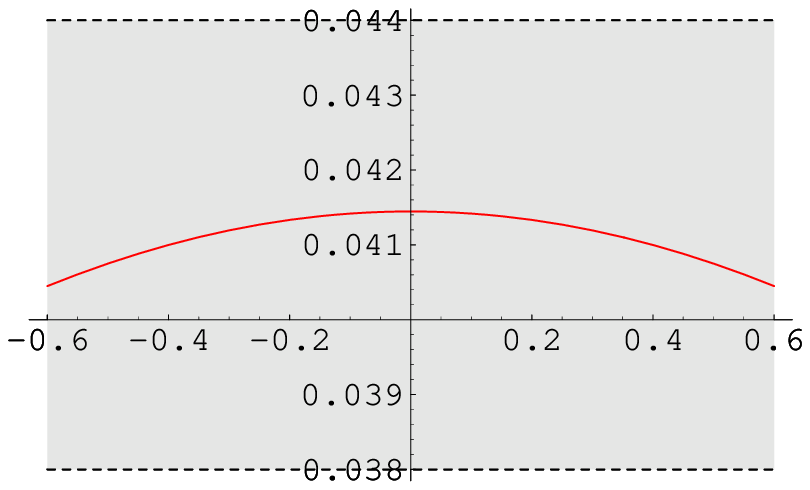}
\end{center}
\begin{center}
\includegraphics[scale=0.5]{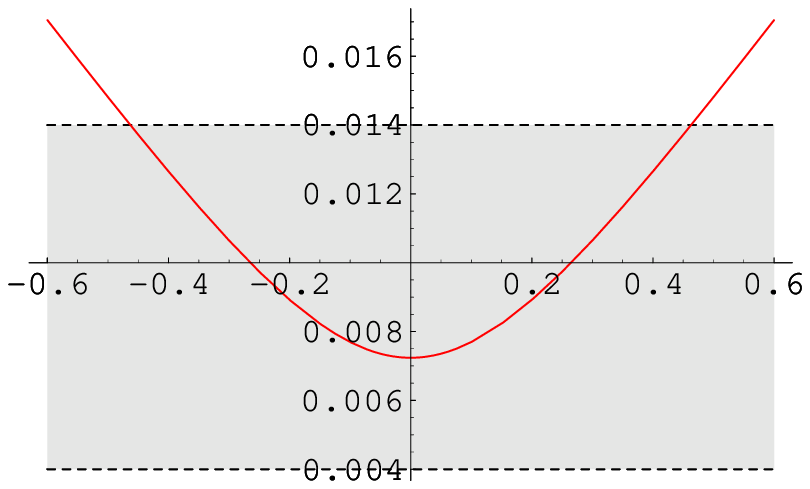}
\includegraphics[scale=0.5]{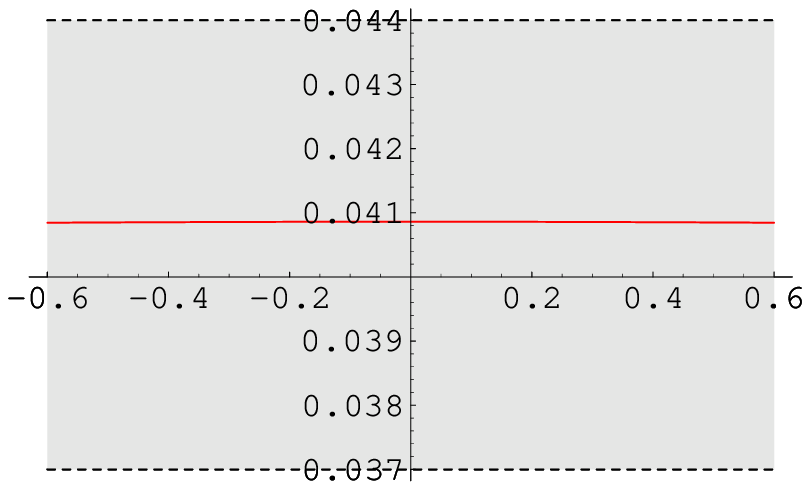}
\includegraphics[scale=0.5]{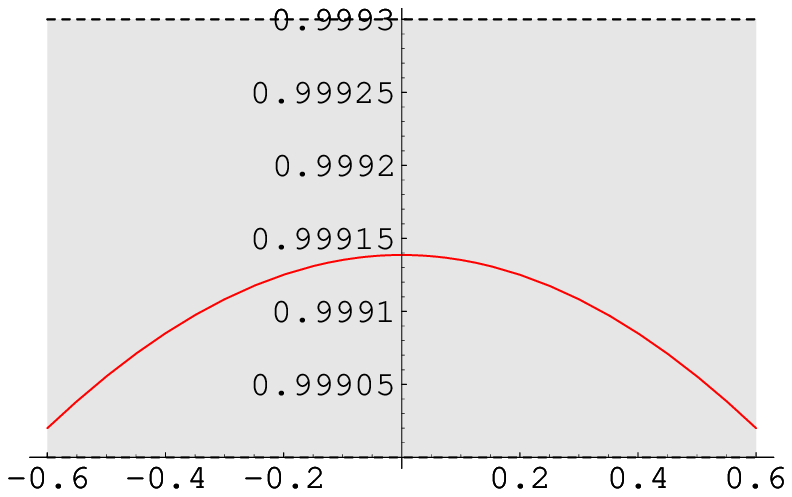}
\end{center} \caption{Variation of $CKM$ entries w.r.t  the parameter $\phi_t$, the position
of each figure overlaps with  its position in the $CKM$
matrix.}\label{fig:ckm-graph-phase-t}
\end{figure}

It is seen from the three cases that the phases have different
contributions on the entries. If we assume that the central values
$(-\gamma_1 ,-\alpha_1,\beta_1, $ $ \beta_2,\alpha_2,\gamma_2 )$
given in eq. (\ref{eq:central-values}) are good values, we can
expand the expression in Eq. (\ref{eq:ckm-th}) in a series around
these values where   $\Delta$'s appear  as fluctuations shown in
the second column of Eq. (\ref{eq:central-values}).

\section{Fluctuations}\label{sec:fluc}

To make the analysis some what easier, we let the complex phases
be initially zero and expand the $U^{CKM}$ around the central
values. If  only the second order terms are selected then we
obtain:

\begin{equation*}\begin{scriptsize}
   V_{r}\approx \left[\begin{array}{ccc}
     1-\frac{1}{2}(\alpha_1-\alpha_2)^2-\frac{1}{2}(\Delta_{\gamma_1}-\Delta_{\gamma_2})^2       & -\Delta_{\gamma_1}+\Delta_{\gamma_2}-\alpha_1 \, \delta     & +\alpha_1-\alpha_2-\Delta_{\gamma_1} \, \delta\\
     \vspace{1mm} \\
     +\Delta_{\gamma_1}-\Delta_{\gamma_2}+\alpha_2 \, \delta       & 1-\frac{1}{2}(\Delta_{\gamma_1}-\Delta_{\gamma_2})^2-\frac{1}{2}\delta^2    & +(\alpha_1-\alpha_2)\Delta_{\gamma_1} + \delta \\
     \vspace{1mm} \\
     -\alpha_1+\alpha_2+\Delta_{\gamma_2} \, \delta     & +(\alpha_2-\alpha_1)\Delta_{\gamma_2} - \delta    & 1-\frac{1}{2}(\alpha_1-\alpha_2)^2-\frac{1}{2}\delta^2 \\
    \end{array}\right]\end{scriptsize}
\end{equation*}
The entries come out  rather interesting. At second order, we see
how the diagonal elements start to differ from each other through
$\delta$. And it is also seen how $\delta$ induces  an asymmetry
between off diagonal terms. Parameterizing the $CKM$ matrix with
respect to the phases will bring changes in the terms. Let us take
the most general case where all phases fluctuate around zero up to
second order, then $V_{\phi_1}$ and $V_{\phi_2}$ will  contribute
to all entries above. The new terms which additively contribute to
each side will be  collected in the following matrix
\begin{equation}\label{ckm-phase}
   V_{c} = \left[\begin{array}{lll}
     V_{11}      & V_{12}     & V_{13}\\
     V_{21}      & V_{22}    & V_{23}\\
     V_{31}      & V_{32}    & V_{33} \\
    \end{array}\right]
\end{equation}
where by definition all  second order terms are collected in $V =
V_{r} + i \, V_{c}$. The terms in the above matrix $V_c$ come out
as
\begin{equation*}
\begin{split}
 V_{11} & = - \frac{1}{2 i }(\phi_c-\phi_s)^2-(\phi_s-\phi_c)\\
 V_{22} & = - \frac{1}{2 i }(\phi_d-\phi_u)^2-(\phi_d-\phi_u)\\
 V_{33} & = - \frac{1}{2 i }(\phi_b-\phi_t)^2-(\phi_b-\phi_t)\\
\end{split}
\end{equation*}
\begin{equation}
\begin{split}
 V_{12} & = +\gamma_2(\phi_c-\phi_s)+\gamma_1(\phi_d-\phi_u)  \\
 V_{21} & = +\gamma_1(\phi_c-\phi_s)+\gamma_2(\phi_d-\phi_u)\\
 V_{13} & = -\alpha_2(\phi_c-\phi_s)+\alpha_1(\phi_b-\phi_t)\\
 V_{31} & = -\alpha_1(\phi_c-\phi_s)+\alpha_2(\phi_b-\phi_t) \\
 V_{23} & = - \delta (\phi_b-\phi_u) \\
 V_{32} & = + \delta (\phi_d-\phi_t) \\
\end{split}
\end{equation}
If each quark in an isospin pair has the same phase then up to
second order  we see from the above expressions that the phase
contributions identically vanish. We have a non-vanishing complex
phase only if the quarks in an isospin pair have different phases.
Then only one unequal phase pair is sufficient to induce $CP$
violation. Using the following definitions:
\begin{equation}
\begin{split}
 \phi_{ud} & = \phi_u-\phi_d \\
 \phi_{cs} & = \phi_c-\phi_s \\
 \phi_{tb} & = \phi_t-\phi_b \\
\end{split}
\end{equation}
the over all amount of  phase $\Phi$  for any choice of
$(-\gamma_1 ,-\alpha_1,\delta, \alpha_2,\gamma_2 )$ could be
calculated from  the determinant of the $CKM$ matrix and comes out
as
\begin{equation}
\begin{split}
\Phi & = e^{-i(\phi_{ud}+ \phi_{cs} + \phi_{tb})} \\
\end{split}
\end{equation}
It is definitely better and more practical to use the exact form
of the $CKM$ matrix given in Eq. (\ref{eq:exact-ckm}) rather than
the parameterized form. Since even at second order, certain
entries deviate from their exact values in a few factors of
$10^{-3}$ and although third order contributions are a good cure
they do  create a mess. The main reason we have parameterized our
expression is to visualize the behavior of the entries. Indeed a
similar behavior is well known from the Wolfenstein
parametrization where the usual $\lambda^3$ , $\rho$ and $\eta$
terms are inevitable.

\section{The Massless Limit}\label{sec:limit1}

It is remarkable that $\beta_1$ and $\beta_2$ take values which
are small deviations around $\frac{\pi}{2}$ and simultaneously
predict acceptable quark masses. Let us  turn the deviations
$\Delta_{\beta_1}$ and $\Delta_{\beta_2}$ temporarily off by
setting them to zero. Recalling the expression in Eq.
(\ref{eq:beta1}) and (\ref{eq:beta2}) we get
\begin{equation}\label{limiting_case}
 \beta_2= \cos^{-1} \left[+\frac{\sqrt{4 \epsilon^2_d a^2_d +k^2_d
   - k_d \sqrt{4 \epsilon^2_d a^2_d+k^2_d}}}{\sqrt{8 \epsilon^2_d a^2_d +2 \, k^2_d}
  }  \right]\rightarrow \frac{\pi}{2}
\end{equation}
The argument of  $\cos^{-1}[\dots]$ becomes zero, which implies
$a_d , k_d \rightarrow 0 $. This is obviously the massless limit.
Note that $k_d$ does not necessarily have to be zero for $\beta_2$
to become $\pi/2$. But it has to be zero so that the third family
receives no mass in the discussed limit. The similar applies to
$\beta_1$ and gives $a_u , k_u \rightarrow 0 $. This is totaly
consistent with the notion of symmetry breaking. The masses result
from a relatively small deviation from an angle. But when we set
$\delta$  identically to zero we see from the expression $V_r$
that still we have non zero entries, in the case that one of the
following asymmetry exists:  $\gamma_1\neq\gamma_2$,
$\alpha_1\neq\alpha_2$, $\phi_u \neq \phi_d$, $\phi_c \neq
\phi_s$, $\phi_b \neq \phi_t$. There is nothing contradictory
about this fact, since in the very  first step of the construction
if $a_u$ , $a_d$ and $k_u$ , $k_d$ are set to zero then the mass
matrices do vanish, and all parameters go symmetric.

\section{The Degenerate Mass Limit}\label{sec:limit2}

The $\pm$ sign convention in the  angles $\beta_1$ and $\beta_2$ is
easily clarified when certain limits  of the parameters in the
simple mass matrices are considered. As a first case we consider
the degenerate mass limit where one has
\begin{equation}
\frac{m_u}{\epsilon_u}= m_c =\frac{m_t}{\epsilon_u} =
\frac{m_d}{\epsilon_d}= m_s =\frac{m_b}{\epsilon_d}
\end{equation}
This spectrum can be achieved by setting $k_u=k_d=0$ and $a_u =
a_d $ $\neq 0$. A second possibility is
\begin{equation}
\frac{m_u}{\epsilon_u}= m_c =\frac{m_t}{\epsilon_u} \neq
\frac{m_d}{\epsilon_d}= m_s =\frac{m_b}{\epsilon_d}
\end{equation}
This degenerate  mass spectrum could be obtained through setting
$k_u=k_d=0$ and $a_u \neq 0 $ $ , a_d\neq0$ and $a_u \neq a_d $.
For the first case we have then
\begin{equation}\label{eq:limiting_case1}
\begin{split}
 \beta_1 & =  \cos^{-1} \left[-\frac{\sqrt{4 \epsilon^2_u a^2_u +k^2_u
   - k_u \sqrt{4 \epsilon^2_u a^2_u+k^2_u}}}{\sqrt{8 \epsilon^2_u a^2_u +2 \, k^2_u}
  } \right] \rightarrow \frac{3 \pi}{4} \\
   \beta_2 & =   \cos^{-1} \left[+\frac{\sqrt{4 \epsilon^2_d a^2_2 +k^2_d
   - k_d \sqrt{4 \epsilon^2_d a^2_2+k^2_d}}}{\sqrt{8 \epsilon^2_d a^2_d +2 \, k^2_d}
  }  \right] \rightarrow \frac{\pi}{4} \\
\end{split}
\end{equation}
and consequently $\delta = \beta_1-\beta_2  \rightarrow \pi/2$ which gives for
the matrix $V_\delta$
\begin{equation}\
 V_{\beta_1} V_{\beta_2}^\dagger = V_\delta =
\left[\begin{array}{ccc}
  0 & 0 &  1  \\
  0 & 1 & 0 \\
  -1  & 0 & 0  \\
\end{array}\right]
\end{equation}
The point of the discussion lies exactly here. If there were no
minus sign in $\beta_1$ , $\delta$ would go to zero and $V_\delta$
would become a unit matrix. The minus sign is essential since when one
lets all angles go symmetric like $\alpha_1=\alpha_2 \neq 0 $,
$\gamma_1=\gamma_2 \neq 0 $, then we still have mixing as should
be expected. If the minus sign were not present in Eq.
(\ref{eq:limiting_case1}), the CKM matrix would have turned into a
unit matrix for $\alpha_1=\alpha_2 \neq 0 $, $\gamma_1=\gamma_2
\neq 0 $. Therefore we choose for $\beta_1$ and $\beta_2$ opposite
signs.

The second case for degenerate quark masses is indeed not that
much a special case. Here $\delta$ will depend on $a_u$ , $a_d$
and other angles as well but the sign convention should be of
course as in the first case.

A final case of interest is when one has $\alpha_1=\alpha_2
=\gamma_2 = \pi/2 $. This decouples the third family from the
first two.

\section{The Nature of the Scaling law}\label{sec:permutations}

For the 3 families $\Psi_1$,$\Psi_2$ and $\Psi_3$, we have
initially introduced the mass matrix
\begin{equation}
M =\left[\begin{array}{ccc}
  k & 0 & a \\
  0 & a & 0 \\
  a & 0 & 0 \\
\end{array}\right]_1
\end{equation}
where for generality the isospin up and down indices are
suppressed. This matrix  produces the scaled masses. It is
possible to construct other matrices with 3 entries of 'a' and one
'k' that  produces the same eigenvalues as well. Now if we impose
a permutation  on the family index such that $\Psi_1 $ is
interchanged with $ \Psi_2 $, keeping $\Psi_3$ untouched we
perform a map on the entries of the mass matrix
\begin{equation}
    \begin{array}{c}
      M_{11} = k \\
      M_{22} = a \\
      M_{31} = a \\
      M_{13} = a \\
    \end{array} \mapsto
    \begin{array}{c}
      M_{22} = k \\
      M_{11} = a \\
      M_{32} = a \\
      M_{23} = a \\
    \end{array}
\end{equation}
one can generate the matrix
\begin{equation}
M = \left[\begin{array}{ccc}
  a & 0 & 0 \\
  0 & k & a \\
  0 & a & 0 \\
\end{array}\right]_2
\end{equation}
which has the same eigenvalues with the initial one ,but is not
identical which means that the mass matrix has no symmetry
property under this permutation. There are four more cases which
are through permutation obtainable
\begin{equation}
\left[\begin{array}{ccc}
  0 & 0 & a \\
  0 & a & 0 \\
  a & 0 & k \\
\end{array}\right]_3\ ,
\left[\begin{array}{ccc}
  a & 0 & 0 \\
  0 & 0 & a \\
  0 & a & k \\
\end{array}\right]_4\ ,
\left[\begin{array}{ccc}
  k & a & 0 \\
  a & 0 & 0 \\
  0 & 0 & a \\
\end{array}\right]_5\  ,
\left[\begin{array}{ccc}
  0 & a & 0 \\
  a & k & 0 \\
  0 & 0 & a \\
\end{array}\right]_6
\end{equation}
These are all the six possible simple mass matrices that produce
the scaled masses. Non of these 6 matrices are identical and the
mass matrix has obviously no permutation symmetry. We have chosen
especially the first one while the the mass eigenvalues are
ordered from low to high i.e., $(m_u,m_c,m_t) $. The point in this
discussion is that the  two mass matrices with label 1 and 3 can
be rotated with matrices of type $V_{\beta}$ to their diagonal
form.  The matrices 2 and 4 can be diagonalized with  $V_{\alpha}$
and the remaining two can be  diagonalized through $V_\gamma$. Any
of the above mass matrices can be used to build the $CKM$ matrix
with same technique to obtain equivalent descriptions. If $k$  is
set in all 6 matrices to zero, we obtain a degenerate mass
spectrum i.e., $(-a,a,a)$, which is not describing our(!) quarks.
Presumably the scaling law is the simplest natural extension of
the degenerate case with the inclusion of the parameter $k$.

\section{Generating the Texture}\label{sec:texture}

The generalized transformations in the flavor space which we based
on the  scaling law, do naturally define a texture. We will look
at the mass matrices given in eq. (\ref{eq:mass-matrices}) and reduce
it to a Texture. The mass matrices are
\begin{equation}
 \begin{split}
   \mathbb{M}^u & =   V_{\gamma_1}^\dagger V_{\alpha_1}^\dagger V_{\phi_1}  M^u V_{\phi_1}^\dagger V_{\alpha_1} V_{\gamma_1}     \\
   \mathbb{M}^d & =  V_{\gamma_2}^\dagger  V_{\alpha_2}^\dagger V_{\phi_2} M^d V_{\phi_2}^\dagger V_{\alpha_2}   V_{\gamma_2}   \\
\end{split}
\end{equation}
First we consider the phaseless case, where all quark phases are
identically set to zero so that $V_{\phi_1}=V_{\phi_2}=I$. Using
the explicit expression for the $V$ matrices the mass matrices
$\mathbb{M}^u$ and $\mathbb{M}^d$ come out as.
\begin{eqnarray}\label{eq:texture}
\left[\begin{array}{ccc}
  F_u & A_u & E_u \\
  A_u & C_u & B_u \\
  E_u & B_u & D_u \\
\end{array}\right] \ \ \ \  , \ \ \ \ \
\left[\begin{array}{ccc}
  F_d & A_d & E_d \\
  A_d & C_d & B_d \\
  E_d & B_d & D_d \\
\end{array}\right]
\end{eqnarray}
where the entries are explicitly
\begin{equation}
\begin{split}
A_u  & = a_u \, s_{\alpha_1} \left( \epsilon_u   s^2_{\gamma_1} - \epsilon^\dagger_u  c^2_{\gamma_1}\right) +\left(k_u-a_u c^2_{\alpha_1} \right)s_{\gamma_1} c_{\gamma_1}  \\
B_u  & = a_u \,  c_{\alpha_1} \left(\epsilon_u s_{\gamma_1} + s_{\alpha_1} c_{\gamma_1}   \right)\\
C_u  & = a_u \, c^2_{\alpha_1} c^2_{\gamma_1} -  \, a_u  s_{2 \gamma_1} s_{\alpha_1} Re[\epsilon_u]+   k_u s^2_{\gamma_1}\\
D_u  & = a_u \, s^2_{\alpha_1} \\
E_u  & = a_u \, c_{\alpha_1} \left( \epsilon_u   c_{\gamma_1}  - s_{\alpha_1} s_{\gamma_1}\right) \\
F_u  & = k_u \, c^2_{\gamma_1} + a_u c^2_{\alpha_1} s^2_{\gamma_1} + a_u s_{\alpha_1} s_{2\gamma_1}Re[\epsilon_u]\\
\end{split}
\end{equation}
\begin{equation}
\begin{split}
A_d  & = a_d \, s_{\alpha_2} \left( \epsilon_d    s^2_{\gamma_2} - \epsilon^\dagger_d  c^2_{\gamma_2}\right) +\left(k_d-a_d c^2_{\alpha_2} \right)s_{\gamma_2} c_{\gamma_2}  \\
B_d  & = a_d \,  c_{\alpha_2} \left(\epsilon_d s_{\gamma_2} + s_{\alpha_2} c_{\gamma_2}   \right)\\
C_d  & = a_d \, c^2_{\alpha_2} c^2_{\gamma_2} -  \, a_d  s_{\gamma_2} s_{2 \alpha_2} Re[\epsilon_d]+   k_d s^2_{\gamma_2}\\
D_d  & = a_d \, s^2_{\alpha_2} \\
E_d  & = a_d \, c_{\alpha_2} \left( \epsilon_d   c_{\gamma_2}  - s_{\alpha_2} s_{\gamma_2}\right) \\
F_d  & = k_d \, c^2_{\gamma_2} + a_d c^2_{\alpha_2} s^2_{\gamma_2} + a_d s_{\alpha_2} s_{2\gamma_2} Re[\epsilon_d]  \\
\end{split}
\end{equation}
Here $s$ and $c$ are shortly for sine and cosine and the subscripts are the arguments.
Since these terms should be real valued, we let $\epsilon_u$ and
$\epsilon_d$ be real quantities. The reason they were defined as
complex variables was to keep track of their conjugation sign. If
we let the phases $V_{\phi_1}$ and $ V_{\phi_2}$ contribute to the
mass matrix we get
\begin{eqnarray}
 \mathbb{M}^u  & =  \left[\begin{array}{ccc}
  \mathcal{F}_u     & \mathcal{A}_u & \mathcal{E}_u \\
  \mathcal{A}^{*}_u & \mathcal{C}_u & \mathcal{B}_u \\
  \mathcal{E}^{*}_u & \mathcal{B}^{*}_u & \mathcal{D}_u \\
\end{array}\right] \ \ \ \   , \ \ \ \ \
 \mathbb{M}^d  & =  \left[\begin{array}{ccc}
  \mathcal{F}_d     & \mathcal{A}_d & \mathcal{E}_d \\
  \mathcal{A}^{*}_d & \mathcal{C}_d & \mathcal{B}_d \\
  \mathcal{E}^{*}_d & \mathcal{B}^{*}_d & \mathcal{D}_d \\
\end{array}\right]
\end{eqnarray}
where the entries are explicitly
\begin{equation}
\begin{split}
\mathcal{A}_u  & = a_u \, s_{\alpha_1} \left( \epsilon_u  e^{i \phi_{tu}}  s^2_{\gamma_1} - \epsilon^\dagger_u e^{-i \phi_{tu}} c^2_{\gamma_1}\right) +\left(k_u-a_u c^2_{\alpha_1} \right)s_{\gamma_1} c_{\gamma_1}  \\
\mathcal{B}_u  & = a_u \,  c_{\alpha_1} \left(\epsilon_u e^{-i \phi_{tu}}s_{\gamma_1} + s_{\alpha_1} c_{\gamma_1}   \right)\\
\mathcal{C}_u  & = a_u \, c^2_{\alpha_1} c^2_{\gamma_1} -  \, a_u   s_{2 \gamma_1} s_{\alpha_1} \frac{1}{2} ( \epsilon_u e^{-i \phi_{tu}}+\epsilon^\dagger_u e^{+i \phi_{tu}})+   k_u s^2_{\gamma_1}\\
\mathcal{D}_u  & = a_u \, s^2_{\alpha_1} \\
\mathcal{E}_u  & = a_u \, c_{\alpha_1} \left( \epsilon_u  e^{-i \phi_{tu}} c_{\gamma_1}  - s_{\alpha_1} s_{\gamma_1}\right) \\
\mathcal{F}_u  & = k_u \, c^2_{\gamma_1} + a_u c^2_{\alpha_1} s^2_{\gamma_1} + a_u s_{\alpha_1} s_{2\gamma_1}\frac{1}{2}( \epsilon_u e^{-i \phi_{tu}}+\epsilon^\dagger_u e^{+i \phi_{tu}})\\
\end{split}
\end{equation}
\begin{equation}
\begin{split}
\mathcal{A}_d  & = a_d \, s_{\alpha_2} \left( \epsilon_d  e^{i \phi_{bd}}  s^2_{\gamma_2} - \epsilon^\dagger_d e^{-i \phi_{bd}} c^2_{\gamma_2}\right) +\left(k_d-a_d c^2_{\alpha_2} \right)s_{\gamma_2} c_{\gamma_2}  \\
\mathcal{B}_d  & = a_d \,  c_{\alpha_2} \left(\epsilon_d e^{-i \phi_{bd}}s_{\gamma_2} + s_{\alpha_2} c_{\gamma_2}   \right)\\
\mathcal{C}_d  & = a_d \, c^2_{\alpha_2} c^2_{\gamma_2} -  \, a_d   s_{2 \gamma_2} s_{\alpha_2} \frac{1}{2} ( \epsilon_d e^{-i \phi_{bd}}+\epsilon^\dagger_d e^{+i \phi_{bd}})+   k_d s^2_{\gamma_2}\\
\mathcal{D}_d  & = a_d \, s^2_{\alpha_2} \\
\mathcal{E}_d  & = a_d \, c_{\alpha_2} \left( \epsilon_d  e^{-i \phi_{bd}} c_{\gamma_2}  - s_{\alpha_2} s_{\gamma_2}\right) \\
\mathcal{F}_d  & = k_d \, c^2_{\gamma_2} + a_d c^2_{\alpha_2} s^2_{\gamma_2} + a_d s_{\alpha_2} s_{2\gamma_2}\frac{1}{2}( \epsilon_d e^{-i \phi_{bd}}+\epsilon^\dagger_d e^{+i \phi_{bd}} )\\
\end{split}
\end{equation}
here  only $\mathcal{D}$  is real  and equals to $D$.  We see that
the model is based on a mass matrix that contains no zeros in its
texture and can be regarded  as  a general Hermitian Matrix
leading to realistic schemes of Mass Matrices as in
\cite{Fritzsch}. In the above expressions it is seen that each
time one phase drops out and we are left with 6 independent
parameters in $\mathbb{M}^u$ and $\mathbb{M}^d$ which are
$k_u,a_u,\alpha_1,\gamma_1,\phi_t,\phi_u$ and are
$k_d,a_d,\alpha_2,\gamma_2,\phi_b,\phi_d$ respectively. One
remarkable thing about $\epsilon_u$ and $\epsilon_d$ is that it
always sticks to the phase $e^{-i\phi_{tu}}$ and $e^{-i\phi_{bd}}$
respectively. The parameters $\epsilon_u$ and $\epsilon_d$ could be absorbed
into the  phase through writing,
\begin{equation}
\epsilon_d e^{-i\phi_{bd}} = e^{-i( \phi_{bd}+\omega_d )} \ \ , \
\ \omega_d = \text{Re}\left[\omega_d\right] + i \,
\text{Im}\left[\omega_d\right] \rightarrow \epsilon_d =
e^{-\text{Im}\left[\omega_d\right]}
\end{equation}
Since $\epsilon_u$ and $\epsilon_d$ should essentially be real, omega
should have no real part. $Re\left[\omega_u\right]=Re\left[\omega_d\right]=0$.
The same could be applied to $\epsilon_u$ as well so that
\begin{equation}
\epsilon_u =
e^{-\text{Im}\left[\omega_u\right]}
\end{equation}

\section{Breaking The Chiral Symmetry}\label{sec:vevs}

In the context of grand unification, It is most natural to set the
Yukawa couplings to
\begin{equation}
   Y_{ij}= \frac{1}{3} \, \left[ \begin{array}{ccc}
      1 & 1 & 1 \\
      1 & 1 & 1 \\
      1 & 1 & 1 \\
    \end{array}\right]
\end{equation}
In any spontaneously broken gauge symmetry, such a Yukawa coupling
would produce only a mass for the third family. Sorting out up
quark and down quark masses into their respective mass matrices,
and  diagonalizing these mass matrices give
\begin{equation}
   \mathcal{M}^u = \left[\begin{array}{ccc}
      0 & 0 & 0 \\
      0 & 0 & 0 \\
      0 & 0 & k_u \\
    \end{array}\right] \ ,  \ \ \ \ \ \
   \mathcal{M}^d = \left[\begin{array}{ccc}
      0 & 0 & 0 \\
      0 & 0 & 0 \\
      0 & 0 & k_d \\
    \end{array}\right]
\end{equation}
One can take $k_u$ and $k_d$ as  the vacuum expectation values of
the Higgs fields. The generation of the masses for the  first and
second  families for the above democratic Yukawa matrices is not
possible. One $could$ let the Yukawa entries depart from unity,
and with a fine tuning it would be possible to fit the current
quark masses and quarks mixing, $but$ there is no predictive power
in such an approach.

The above diagonal mass matrices are  derivable from the simple
mass matrices given in Eq.(\ref{eq:smm}) through taking
$\beta_1=\pi/2$ and $\beta_2=\pi/2$ with $a_u = a_d = 0$. Which
connects the above mass matrices with those of the model presented
here. The generation of non-zero values of $a_u $ and $ a_d $ , in
the framework of GUT's could be interesting.

In the limit of $N_f$ massless quarks , the $QCD$ lagrangian has a
well known exact global chiral symmetry $G_{LR}$ = $SU(N_f)_L
\times SU(N_f)_R $ which acts on the left and right handed quarks.
If we consider that the spontaneous breakdown of the gauge
symmetry is accompanied by a spontaneous breakdown of the chiral
symmetry in the QCD sector~\cite{Nambu-Chiral1}
~\cite{Nambu-Chiral2}, it would be possible to introduce the $a_u
$ and $ a_d $ terms. These terms are relatively small. From the
known spectrum of quark masses we have:
\begin{equation}\label{eq:hierarchy}
\begin{split}
    & \frac{a_u}{k_u} \approx \frac{1}{281} \\
\end{split}
\begin{split} \ \ ,  \ \ \ \
    & \frac{a_d}{k_d} \approx \frac{1}{27} \\
\end{split}
\end{equation}
The vevs  $a_u $ and $ a_d $ give then $degenerate$ masses to the
quarks, and  with the inclusion of $k_u$ and $k_d$, the
hierarchical mass spectrum could be recovered as given in Eq.
(\ref{eq:scaling-law}) with the mass matrices:
\begin{equation}
   M^u = \left[\begin{array}{ccc}
      k_u & 0 & a_u \\
      0 & a_u & 0 \\
      a_u & 0 & 0 \\
    \end{array}\right] \ ,  \ \ \ \ \ \
   M^d = \left[\begin{array}{ccc}
      k_d & 0 & a_d \\
      0 & a_d & 0 \\
      a_d & 0 & 0 \\
    \end{array}\right]
\end{equation}
It is seen from Eq. (\ref{eq:ka1})  that if $k_u
>> a_u$ and  $k_d>> a_d$, the chiral breakdown has no significant
contribution to the bottom and top quark masses. The prediction of
the 6 quark masses is  reduced to determining $k_u, a_u$ and $k_d
, a_d$. It is also of interest whether the parameters $\epsilon_u$
and $\epsilon_d$ could be obtained from radiative correction.

The strong CP violation and $\theta$ problem is related with the
nature of the higgs sector~\cite{Ross}.i.e., the higgs fields
giving masses to up and down quarks should not be related over
conjugation so that the strong CP phase can be naturally
moderated.

An $SO(10)$ model with the higgs fields of $126$ and $10$ namely
with the submultiplets $(2,2,15)$ and $(2,2,1)$ respectively,
could present a rich framework for handling these problems.  In
such a model the mass eigenstates of the gauge bosons and various
mixing angles will depend on the vacuum expectation values of the
higgs fields which $also$ determine $k_u$ and $k_d$. We defer a
detailed analysis to a separate work.

\section{Mass Inversion : $ m_d > m_u$}\label{sec:mass-inv}

A final point we discuss is the well known observation that u-type
quarks are heavier than the d-type quarks except for $ m_d > m_u$.
The simple mass matrices have intrinsically a nice structure which
under certain conditions can give rise to such an inversion in the
mass spectrum. The relevant case is to consider all the range in
which $k_u> k_d$ and $a_u> a_d$. Then from the eigenvalues given
in Eq. (\ref{eq:mass_eigenstates}) we always have $m_t> m_b$ and
$m_c> m_s$. In this range we see that it is possible to have both
situations namely,  $m_d>m_u$ or $m_d<m_u$ for ceratin values of
the parameters. We will not try to figure out the conditions, but
we find it in particular interesting to point out that an
inversion is possible for certain values of the parameters under
the condition $k_u > k_d$ and $a_u > a_d$ which obviously dictates
a mass hierarchy among down and up type quark masses.

\section{Conclusion}\label{sec:conclusion}

The model presented above serves to fill a missing gap between the
$CKM$ matrix and the quark masses in many respects. It describes
a non-conventional way to build the $CKM$ matrix. We started with
the assumption that quark masses obey a scaling law, and extended
the construction on general rotations in flavor space. The
parameters of the rotation describe deviations from  an initially
symmetric condition, which is completely compatible with the idea
of symmetry breaking. The results have a mutual character. First
of all it allows to determine quark masses from the experimentally
obtained $CKM$ entries over the angle $\delta=\beta_1-\beta_2$
where $\beta_1$ and $\beta_2$ are related to the parameters $a_u$
, $k_u$ and $a_d$ , $k_d$. It can also be used reversely such that
quark masses can directly influence our knowledge on the $CKM$
entries.

In our model the mass matrix is not based on arbitrary textures
but such that the  initial mass matrices $M^u$ and $M^d$ generate
the simple scaling law among quark masses, regardless of the
values of $a_u$ , $k_u$ and $a_d$, $k_d$. It is then natural to
assume  that the  mass matrices we started with were not $M^u$ and
$M^d$ but,
\begin{equation*}
\begin{split}
 &  \left( V_{\gamma_1}^\dagger V_{\alpha_1}^\dagger V_{\phi_1}  M^u
 V_{\phi_1}^\dagger V_{\alpha_1} V_{\gamma_1} \right) \\
 &  \left( V_{\gamma_2}^\dagger V_{\alpha_2}^\dagger V_{\phi_2}  M^d
 V_{\phi_2}^\dagger V_{\alpha_2} V_{\gamma_2}\right) \\
\end{split}
\end{equation*}
respectively  which are subject to diagonalization as in Eq.
(\ref{eq:f}). It seen that from the structural point of view that
these mass matrices can be classified as non-zero textures and are
quite general expressions. The simple $M^u$ and $M^d$ mass
matrices are then initially containing the information of the
magnitude of the masses $solely$, but not the complete information
of the eigenstates, which in the model is  achieved trough the
rotations.

We have given a series expansion of the $CKM$ matrix which is
capable of explaining at second order  how various entries differ
from each other. It is also nice to see that a slight difference
in the way we parameterize the $CKM$ matrix does not really matter
and can even be extremely predictive. Finally we would like to
admit that the model can predict each $CKM$ entry within the
currently accepted values.

The scaling law might be  consistent with "quark masses" and
"quark-mixing". The success in the prediction of the $CKM$ entries
also might give an end to $Texture$ hunting as we discussed in
some detail.

\bibliographystyle{unsrt}
\bibliography{main}

\end{document}